\documentclass[preprint,NumberedRefs]{JASA}

\usepackage{algpseudocode}

\begin{document}
\nolinenumbers
\title{Equivalence between angular spectrum-based and multipole expansion-based formulas of the acoustic radiation force and torque}

\author{Zhixiong Gong}
\email{Corresponding author: zhixiong.gong@iemn.fr}
\affiliation{Univ. Lille, CNRS, Centrale Lille, Yncréa ISEN, Univ. Polytechnique Hauts-de-France, UMR 8520 - IEMN, F- 59000 Lille, France}

\author{Michael Baudoin}
\affiliation{Univ. Lille, CNRS, Centrale Lille, Yncréa ISEN, Univ. Polytechnique Hauts-de-France, UMR 8520 - IEMN, F- 59000 Lille, France}
\affiliation{Institut Universitaire de France, 1 rue Descartes, 75005 Paris}
\preprint{Gong $\&$ Baudoin, JASA}		%  if you want want this message to appear in upper right corner of title page

\date{\today} 

\begin{abstract}
\nolinenumbers
Two main methods have been proposed to derive the acoustical radiation force and torque applied by an arbitrary acoustic field on a particle: The first one relies on the plane wave angular spectrum decomposition of the incident field (see [Sapozhnikov \& Bailey, J. Acoust. Soc. Am. 133, 661–676 (2013)] for the force and [Gong \& Baudoin, J. Acoust. Soc. Am. 148, 3131–3140 (2020)] for the torque), while the second one relies on the decomposition of the incident field into a sum of spherical waves, the so-called multipole expansion (see [Silva, J. Acoust. Soc. Am. 130, 3541–3544 (2011)] and [Baresh \textit{et al.}, J. Acoust. Soc. Am. 133, 25–36 (2013)] for the force,  and  [Silva \textit{et al.}, EPL 97, 54003 (2012)] and [Gong \textit{et al.}, Phys. Rev. Applied 11, 064022 (2019)] for the torque). In this paper, we formally establish the equivalence between the expressions  obtained with these two methods for both the force and torque.
\end{abstract}

% PACS, the Physics and Astronomy
                                   % Classification Scheme.
%\keywords{Suggested keywords}%Use showkeys class option if keyword display desired

\maketitle
\nolinenumbers
%--------------------------------------
\section{\label{sec:introduction}Introduction}

Since the seminal works of Rayleigh \cite{pm_rayleigh_1902,pm_rayleigh_1905}, Langevin \cite{ra_biquard_1932a,ra_biquard_1932b} and Brillouin \cite{jpr_brillouin_1925,ap_brillouin_1925}, many expressions of the acoustic radiation force and torque applied by various acoustic fields on different types of particle have been derived. King \cite{king1934acoustic} was the first to propose an expression of the acoustic radiation force applied on a rigid sphere by a plane (standing or progressive) wave. This expression was extended later on by Yosika \& Kawasima \cite{a_yosika_1955} for compressible particle and  Hasegawa \& Yiosika \cite{jasa_hasegawa_1969} for an elastic sphere. The case of spherical and focused incident waves was addressed by Embleton \cite{jasa_embleton_1954} and Chen \& Apfel \cite{jasa_chen_1996} for rigid and elastic spheres respectively. Nevertheless all these cases assume axisymmetric incident fields centered on the particle, which considerably simplifies the problem and does not enable to compute the 3D trapping force applied by a selective tweezer on an object \cite{baudoin2020acoustic}. The case of arbitrary acoustic field was at this point only treated in the framework of the Long Wavelength Regime (LWR), i.e. for particle much smaller than the wavelength \cite{spd_gorkov_1962}. Concerning the torque, the very existence of a torque applied on a spherical particle requires the existence of a momentum carried out by the wave, which cannot be obtained with an axisymmetric acoustic field. Busse \& Wang \cite{jasa_busse_1981} demonstrated the role played by the viscous boundary layer on the Torque applied by orthogonal acoustic waves on a spherical particle in the LWR. Later on, Zhang \& Marston \cite{zhang2011angular} proposed an expression of the axial acoustic radiation torque acting on an axisymmetric particle centered on the axis of a cylindrical acoustical vortex. But again the proposed expressions assume certain symmetry of the incident beam and specific location of the scatterer.

The treatment of the general problem of the acoustic radiation force and torque applied on a spherical particle of arbitrary size requires to solve three major issues: First, the incident field must be decomposed into a sum of elementary waves suitable for the treatment of the scattering problem and then the caculation of the force and torque. In the angular spectrum method (ASM) \cite{sapozhnikov2013radiation,gong2020ART}, the incident field is decomposed into a sum of plane wave assuming the prior knowledge of the incident field in a source plane. In the multipole expansion method (MEM)  the incident field is decomposed into a sum of spherical waves \cite{silva2011expression,silva2012radiation,baresch2013three,gong2019t,gong2019reversals}, whose respective contribution (the beam shape coefficients) can be calculated by different methods \cite{zhao2019computation}. Second, the scattering problem must be solved. For an arbitrary wave this task is complexified by the non axisymmetry of the incident acoustic field. The angular spectrum method alleviates the problem by using the fact that the  solution of the scattering problem for a plane wave is known. Nevertheless, each plane wave of the angular spectrum decomposition have a different incident angle. This problem was solved by Sapozhnikov and Bailey \cite{sapozhnikov2013radiation} by using the Legendre addition theorem. In the multipole expansion method, the scattering problem was solved for an arbitrary spherical wave. It was shown by Baresch et al. \cite{baresch2013three} that the problem degenerates to the one of an incident plane wave so that the classical scattering coefficients can be used (see appendix A in ref \citep{baresch2013three}). Third, the force and torque can be calculated by integrating the time-averaged linear and angular radiation stress tensor over the particle surface, respectively. Such integration over the particle surface can be tedious to perform directly since (i) the particle surface is vibrating and hence is varying over time, (ii) the particle geometry may be complex in the case of non-spherical particles and (iii) the existence of viscous and thermal boundary layers must be considered in the near field. It was first shown by Brillouin \cite{jpr_brillouin_1925,ap_brillouin_1925}, that the integral over the vibrating surface of the particle can be transferred to a still surface by replacing the stress tensor by the so-called Brillouin tensor, which includes the flux of momentum through this steady surface. Later on, it was shown that the integral can be transferred to a closed surface in the far field by using the momentum \cite{westervelt1951theory,westervelt1957acoustic} and angular momentum balances \cite{maidanik1958torques,jasa_fan_2008,jasa_zhang_2011} in the surrounding fluid and the Gauss divergence theorem \cite{westervelt1951theory} or the Reynolds transport theorem \cite{jasj_hasegawa_2000}. Hence, by choosing a spherical surface in the far field,  (i) the integration is conducted over a simpler surface (concentric with the particle center) and (ii) the far-field approximation enables to use asymptotic expressions for Bessel and Hankel functions which simplifies the integration procedure. This also enables the treatment of non spherical particles, e.g. using T-matrix method \cite{gong2019reversals}.

Of course, the values of the acoustic radiation force and torque must be independent of the method used to calculate them. While some links between some of the expressions of the ARF available in the literature have been previously evoked \cite{thomas2017acoustical,baudoin2020acoustic}, there is no explicit demonstration of the link between these complex formula. The present paper aims at clarifying this point and establishing formally the equivalence between the different expressions of the ARF and ART derived with different approaches.

%---------------------------------------------------------------------------------
\section{\label{sec:relation between Hnm and Anm} Decomposition of the incident field}

In the multipole expansion method (MEM) \cite{silva2011expression,silva2012radiation,baresch2013three,gong2019reversals,gong2019t}, the incident acoustic potential is directly decomposed in the spherical waves basis as follows:
\begin{equation}
\Phi_{i} =\Phi_{0} \sum_{n=0}^{\infty} \sum_{m=-n}^{n} a_{n}^m j_{n}(k r) Y_{n}^m(\theta, \varphi) e^{-i \omega t}, \label{incident potential MEM}
\end{equation}
with $a_n^m$ the incident \textit{beam-shape coefficients} (BSC), which sets the weight of each spherical waves, $Y_{n}^m(\theta, \varphi)$ the normalized spherical harmonics defined by:
\begin{equation}
Y_{n}^m(\theta, \varphi)= \sqrt{\frac{2 n+1}{4 \pi} \frac{(n-m) !}{(n+m) !}} P_{n}^m(\cos \theta) e^{i m \varphi},
\label{spherical harmoncis Ynm}
\end{equation}
with $(r,\theta,\varphi)$ the spherical coordinates, $\Phi_0$ the potential amplitude, $j_n$ the Bessel function of the first kind, $k$ the wavenumber, and $P_{n}^m$ the associated Legendre functions. Note that only the Bessel functions of the first kind appear in this expression since the incident field exists in absence of the scatterer and hence must be finite in  $(r=0)$, hence eliminating the Bessel functions of the second kind which are singular at this point. A general method to determine the beam shape coefficients (inspired by previous work in optics) for an arbitrary field was introduced by Baresch et al. \cite{baresch2013three} and various methods were tested and compared by Zhao et al. \cite{zhao2019computation}.

In the angular spectrum method (ASM) \cite{sapozhnikov2013radiation,gong2020ART}, the calculation starts from the prior knowledge of the incident pressure field in a source plane ($z=0$) $p_i |_{z=0} = p_i(x,y,0)$, and its decomposition into a sum of plane waves:
\begin{equation}
p_i(x,y,z) = \frac{1}{4 \pi^2} \\ \times \iint_{k_x^2 + k_y^2 \leq k^2} S(k_x,k_y) e^{i k_{x} x + i k_{y} y + i \sqrt{k^2  - k_x^2 - k_y^2} z}d k_{x} d k_{y}.
\end{equation}
using the angular spectrum decomposition (2D spatial Fourier transform) of the source plane field:
\begin{equation}
\begin{aligned}
S\left(k_{x}, k_{y}\right)=\int_{-\infty}^{+\infty} \int_{-\infty}^{+\infty}  p_{i}(x, y, 0) e^{-i k_{x} x - i k_{y} y} d x d y
\end{aligned},
\label{S(kx,ky)}
\end{equation}
with $k_x$ and $k_y$ are the lateral components of the wavenumber $\mathbf{k}$ in Cartesian coordinates, $k_z^2 = k^2  - k_x^2 - k_y^2$ and $k = \omega / c$. 

If the angle $\gamma$ between the position vector $\mathbf{r} = x \mathbf{x} + y \mathbf{y} + z \mathbf{z}$ and the wavevector $\mathbf{k}$ is introduced, we see clearly that the incident field is nothing but the sum of plane waves $p_i^\mathbf{k}$ with different incident angles $\gamma$:
$$
p_i^\mathbf{k}(x,y,z) = S(k_x,k_y) e^{i k r \cos(\gamma)}
$$
Using (i) the known decomposition of a plane wave with an incident angle $\gamma$ into spherical waves and (ii) the Legendre addition theorem to express the final result as a function of the absolute spherical coordinate $(\theta, \varphi)$ instead of the auxiliary angle $\gamma$, Sapozhnikov \& Bailey \cite{sapozhnikov2013radiation} were able to express the incident field into a sum of spherical waves:
\begin{equation}
p_{i}=\frac{1}{\pi} \sum_{n=0}^{\infty} \sum_{m=-n}^{n} i^{n} H_{nm} j_{n}(k r) Y_{n}^m(\theta, \varphi), \label{pressure ASM inc}
\end{equation}
with the ASM-based BSC $H_{nm}$ describing the respective weight of each spherical wave:
\begin{equation}
H_{n m}=\iint_{k_{x}^{2}+k_{y}^{2} \leq k^{2}} S\left(k_{x}, k_{y}\right) [Y_{n }^{m}\left(\theta_{k}, \varphi_{k}\right)]^* d k_{x} d k_{y},
\label{H_nm}
\end{equation}
The asterisk $*$ designates the complex conjugate, and the angle parameters $(\theta_{k}, \varphi_{k})$ in the Fourier space have the relation: $\cos\theta_k = [1-\left(k_{x}^{2}+k_{y}^{2}\right) / k^{2}]^{1/2}$ and $\varphi_k = \arctan \left(k_{y} / k_{x}\right)$. 

This decomposition into a sum of spherical waves is necessary to compute the force and torque since the total field (incident + scattered) needs to be integrated over an arbitrary closed surface surrounding the particle, which for commodity will be chosen as a spherical surface in the far field as discussed in section \ref{sec:calculation of the Force and Torque}. The $H_{nm}$ coefficients can be easily obtained when the field is known in a source plane by using the Spatial Fast Fourier Transform of the incident field, which makes the ASM method very convenient to compute the force applied on a particle by a field generated by a planar transducer \cite{pre_jimenez_2016,prap_riaud_2017,apl_jimenez_2018,sa_baudoin_2019,baudoin2020naturecell}.

The comparison of Eqs. (\ref{incident potential MEM}) and (\ref{pressure ASM inc}) and use of the relationship between the velocity potential and pressure $p_{i}=i \omega \rho_{0} \Phi_{i}$ (with $\omega$ the angular frequency and $\rho_0$ the fluid density), enables to establish the relationship between the incident BSC $a_{n}^m$ and the angular-spectrum based BSC $H_{nm}$
\begin{equation}
a_{n}^m=\frac{1}{\pi \omega \rho_{0} \Phi_{0}} i^{n-1} H_{n m}.
\label{a_nm & H_nm}
\end{equation}
which is essential to prove the equivalence of MEM and ASM based ARF and ART formulas. Note that an equivalent form of Eq. (\ref{a_nm & H_nm}) has been given in Eq. (15) of Ref.\cite{zhao2019computation} by comparing two expressions of acoustic pressure.

\section{\label{sec:scattering} Resolution of the scattering problem}

In the MEM, the scattered field, as the incident field is decomposed directly into a sum of spherical waves:
\begin{equation}
\Phi_{s} = \Phi_{0} \sum_{n=0}^{\infty} \sum_{m=-n}^{n} s_{n}^m h_{n}^{(1)}(k r) Y_{n}^m(\theta, \varphi) e^{-i \omega t}, \label{scattered potential MEM}
\end{equation}
with $s_n^m$ the beam shape coefficient of the scattered field. Note that this time the scattered field is expressed in terms of the Hankel function of the first kind since the scattered field is an outgoing wave, hence eliminating the Hankel function of the second kind (corresponding to converging wave in the convention used here for the temporal part of the wave $e^{-i \omega t}$). The expression of the scattered beam shape coefficients as a function of the incident beam shape coefficients requires to solve the scattering problem, i.e. to determine the \textit{partial wave coefficients} $A_{n}^m$ defined by $s_{n}^m = A_{n}^m a_{n}^m$. These coefficients depend on the particle shape, material composition and surface boundary condition. In the MEM, the solution of the scattering problem is \textit{a priori} not known since the axisymmetry and resulting simplifications used in the case of plane waves can no longer be invoked. The complete problem was solved by Baresh et al. \cite{baresch2013three} for an elastic sphere through the introduction of three scalar potentials, (one for the longitudinal wave and the two Debye potential for the shear wave, solutions of the wave equation and then applying the boundary conditions). It was shown that in fact the problem degenerates to the one of plane incident wave, so that the partial waves coefficient $A_n$ computed for a plane wave, which do not depend on the index $m$,  can be used. Note that in this simplified case, people sometime introduce the so-called \textit{scattering coefficients} $S_n$  linked to the partial wave coefficients by the formula  $A_n = (S_n - 1)/2$. Note also that in the general case of nonspherical particles, the partial wave coefficients can be determined using the transition matrix method \cite{gong2019t,gong2019reversals,gong2018thesis} which makes the theory operable for nonspherical shapes, such as spheroids \cite{gong2016arbitrary} and finite cylinders \cite{gong2017t,gong2017analysis}.

In the angular spectrum method, the treatment relies on known results for the scattering of a plane wave by a sphere. Indeed, (i) the incident field has been decomposed into a sum of plane waves and (ii) the solution of the scattering problem is known for each plane wave. Hence using these solution for each plane wave and then using (i) the decomposition of a plane wave into a sum of spherical waves and (ii) the Legendre addition theorem, the scattered field can also be decomposed into a sum of spherical waves:
\begin{equation}
p_{s}=\frac{1}{\pi} \sum_{n=0}^{\infty} \sum_{m=-n}^{n} i^{n} H_{n m} A_{n}^m h_{n}^{(1)}(k r) Y_{n}^m (\theta, \varphi), \label{scattered pressure ASM}
\end{equation}

\section{\label{sec:calculation of the Force and Torque} Calculation of the force and torque}

The last step, which is common to ASM and MEM is to compute the integral of the stress tensor or angular stress tensor over the surface of the particle to compute the force and torque, respectively. One major difficulty comes from the fact that the surface of the particle is vibrating. This problem can be overcome in two ways: firstly, using Lagrangian coordinates instead of Eulerian coordinates, and secondly, transferring the integral to a still surface by substracting the flux of momentum (flux of angular momentum) to the stress tensor (angular stress tensor) for the force and torque respectively as first demonstrated by Brillouin (for the Force) \cite{ap_brillouin_1925,jpr_brillouin_1925}. To simplify the calculation, these integrals can be transported to any surface surrounding the particle, e.g. for simplicity a spherical surface in the far field as demonstrated by Westervelt for the force \cite{westervelt1951theory,westervelt1957acoustic} and Maidanik and others for the torque \cite{maidanik1958torques,jasa_fan_2008,jasa_zhang_2011}. Using these results, the integrals to compute the force $\mathbf{F}$ and torque $\mathbf{T}$ can be written under the following form in terms of the acoustic potential $(\Phi_{i,s})$ of the incident and scattered field as:
\begin{eqnarray}
\mathbf{F} & = & \frac{\rho_{0} k^{2}}{2}  \iint_{S_{0}} \operatorname{Re}\left\{\left(\frac{i}{k} \frac{\partial \Phi_{i}}{\partial r}-\Phi_{i}\right) \Phi_{s}^{*}-\Phi_{s} \Phi_{s}^{*}\right\} \mathbf{n} d S,
\label{ARF based on acoustic potential} \\
\mathbf{T} & = &\frac{\rho_{0}}{2} \operatorname{Im}\left\{\iint_{S_{0}}\left(\frac{\partial \Phi_{i}^{*}}{\partial r} \mathbf{L} \Phi_{s}+\frac{\partial \Phi_{s}^{*}}{\partial r} \mathbf{L} \Phi_{i}+\frac{\partial \Phi_{s}^{*}}{\partial r} \mathbf{L} \Phi_{s}\right) d S\right\},
\label{torque in potential} 
\end{eqnarray}
where $S_0$ is a closed spherical surface in the far field centered at the mass center of the particle,
$\rho_0$ is the density at rest, `Re'' means the real part of a complex number, ``Im'' designates the imaginary part, $\textbf{n}$ is the outward unit normal vector, and the differential surface area is $d S=r^{2} \sin \theta d \theta d \varphi$ with $\theta$ and $\varphi$ the polar and azimuthal angles, $\mathbf{L}=-i(\mathbf{r} \times \nabla)$ is the angular momentum operator, with its components in the three directions $L_{x,y,z}$ and the recursion relations of the normalized spherical harmonics with ladder operators $L_{\pm}$ given in detail in Appendix \ref{Appendix D}.

In the next section we establish the link between the different formulas obtain in the literature.

\section{\label{arf equivalence} Equivalence of the three-dimension ARF formulas}

Expressions of the ARF exerted by an arbitrary field on an arbitray located spherical scatterer has been established independently by 3 different groups: Silva \cite{silva2011expression} and Baresch \textit{et al.} \cite{baresch2013three} with a MEM, and Sapozhnikov \& Bailey based on the ASM \cite{sapozhnikov2013radiation}. The equivalence between the formulas obtained by Baresch \textit{et al.} \cite{baresch2013three} and Sapozhnikov \& Bailey \cite{sapozhnikov2013radiation} has been briefly discussed by Thomas et al.\cite{thomas2017acoustical,baudoin2020acoustic}, while the equivalence with Silva's work has not been investigated yet.  In this section, the reason for the different forms of ARF formulas by Silva \cite{silva2011expression} and Baresch \textit{et al.} \cite{baresch2013three} is provided (since both use the MEM), while pointing out some minor existing issues in the formula and at the same time, for the first time, providing detailed proof of the equivalence of the ARF formulas for the three work.

\subsection{Equivalence between MEM formula and compact formulation}

\subsubsection{\label{Silva's ARF} MEM formula by Silva and Gong \textit{et al.} and reindexing}

Following the work of Silva \cite{silva2011expression} and of Gong \textit{et al.} \cite{gong2019t}, the dimensionless ARF formulas in terms of the incident $a_n^m$ and scattered $s_n^m$ BSC are obtained by substituting Eq. (\ref{incident potential MEM}) and (\ref{scattered potential MEM}) into Eq. (\ref{ARF based on acoustic potential}) and conducting several algebraic calculations given in Eqs. (11-13) of Ref. \cite{silva2011expression} by Silva or Eqs. (12-14) of Ref. \cite{gong2019t} by Gong \textit{et al.}. The ARF formulas can be therefore obtained based on the relation with the dimensionless ARF according to Eq. (10) in Ref. \cite{gong2019t}.
Note that for the two separate derivations, different asymptotic expressions of velocity potentials in the far-field are used: Silva uses trigonometric functions \cite{silva2011expression} [see Eq. (4) in his paper], while Gong \textit{et al.} use the exponential functions \cite{gong2019t} to approximate the Bessel and  Hankel functions.
In addition, Gong \textit{et al.}'s work is an extension of numerical implementation for non-spherical shapes by using the T-matrix method \cite{gong2019t}.

However, the ARF formulas by Silva \cite{silva2011expression} and Gong \textit{et al.} \cite{gong2019t} missed a re-indexing step in the scattered BSC ($s_{n-1}^{m+1}$, $s_{n-1}^{m-1}$ and $s_n^{-m-1}$), as pointed out recently \cite{thomas2017acoustical,baudoin2020acoustic}. Here, we explicit the reason for the index issue and provide the good expressions: Silva and Gong \textit{et al.} use the simplified double summation symbol $\sum_{nm}$ to represent $\sum_{n=0}^{\infty} \sum_{m=-n}^{n}$ to conduct the integral process involving the product of two spherical harmonics [see Eq. (11) in Ref. \cite{gong2019t}].
A mistake appears since the regime of $m$ should be correctly chosen for the spherical harmonics $Y_{n-1}^{m\pm1}$ and $Y_{n-1}^{m}$ [as given in Eqs. (\ref{cos Ynm}) and (\ref{exp Ynm})] based on the definition of spherical harmonics $Y_n^m$ with $|m| \leq n$, which means  $\sum_{nm}$ is not always $\sum_{n=0}^{\infty} \sum_{m=-n}^{n}$.

In this work, we re-derive the formulas following the right indexes ($n,m$) and therefore get the correct forms as (see details in Appendix \ref{Appendix B})
\begin{subequations}
\begin{eqnarray}
F_{x} = \frac{\rho_{0} \Phi_{0}^{2}}{4}  \operatorname{Im} \Bigg\{ \sum_{n=0}^{\infty} \sum_{m=-n}^{n} &\bigg[&
b_{n+1}^{-m} [\left( a_{n}^m + s_{n}^m \right) s_{n+1}^{m-1*} - \left( a_{n+1}^{m-1} + s_{n+1}^{m-1} \right) s_{n}^{m*}] \label{Fx gong baudoin a} \\
& + &  b_{n+1}^{m} [\left( a_{n+1}^{m+1} + s_{n+1}^{m+1} \right) s_{n m}^* - \left( a_{n}^m + s_{n}^m \right) s_{n+1}^{m+1*}] \bigg] \Bigg\}
,  \nonumber \\
F_{y}=\frac{\rho_{0} \Phi_{0}^{2}}{4} \operatorname{Re} \Bigg\{ \sum_{n=0}^{\infty} \sum_{m=-n}^{n} & \bigg[ &
b_{n+1}^{-m} \left[\left( a_{n}^m + s_{n}^m \right) s_{n+1}^{m-1*} + \left( a_{n+1}^{m-1} + s_{n+1}^{m-1} \right) s_{n}^{m*} \right] \label{Fy gong baudoin b} \\
& + &  b_{n+1}^{m} [\left( a_{n+1}^{m+1}+s_{n+1}^{m+1} \right) s_{n}^{m*} + \left( a_{n}^m + s_{n}^m \right) s_{n+1}^{m+1*}]
\bigg]\Bigg\}, \nonumber  \\
F_{z} = \frac{\rho_{0} \Phi_{0}^{2}}{2}  \operatorname{Im} \Bigg\{ \sum_{n=0}^{\infty} \sum_{m=-n}^{n} & c_{n+1}^{m} & \left[ \left( a_{n+1}^{m*}+s_{n+1}^{m*} \right) s_{n}^m +\left( a_{n}^m +s_{n}^m \right) s_{n+1}^{m*} \right] \Bigg\}. \label{Fz gong baudoin c}
\end{eqnarray}
\label{ARF gong baudoin}
\end{subequations}
where $n \in [0,\infty]$ and $m \in [-n,n]$, and the coefficients $b_n^m$ and $c_n^m$ defined in terms of $n$ and $m$ are given in the Appendix \ref{Appendix A}.

Note that despite the index issue, the numerical computations in Ref.\cite{gong2019t} are correct since the erroneous additional terms were cancelled in the numerical procedure. This can be further verified by the comparison of results by Gong \textit{et al.} with the partial wave based results by Marston \cite{marston2006axial}. Note also that this set of formulas can be written in a much more compact form using the relation $s_n^m = A_n^m a_n^m$ which will be given in Sec. \ref{Baresch et al's ARF}.

%-----------------------------------------------------------------
\subsubsection{\label{Baresch et al's ARF} The ARF formulas by Baresh et al.}
Another set of ARF formulas based on the MEM, was derived by Baresh, Thomas \& Marchiano for 3D ARF on an arbitrarily located elastic sphere, as given by Eqs. (14-16) in Ref. \cite{baresch2013three} [reorganized as Eqs. (1-3) by Zhao, Thomas \& Marchiano in Ref. \cite{zhao2019computation}]. Note that there is a typo for the regime of index $m$ \cite{baresch2013three,zhao2019computation,baudoin2020acoustic}: it should be $|m| \leq n$ instead of $|m|<n$ (otherwise the formulas are not equivalent to those by Sapozhnikov \& Bailey \cite{sapozhnikov2013radiation}). The ARF formulas with the right index regimes by Thomas and colleagues are equivalent to the above corrected version (see Eq. \ref{ARF gong baudoin}) of Silva and Gong \textit{et al.}'s formulas.

The difference between formulas by Silva \cite{silva2011expression} (or Gong \textit{et al.} \cite{gong2019t}) and Baresh, Thomas \& Marchiano \cite{baresch2013three} are the following: 
(i) Silva uses the incident $a_{n}^m$ and scattered $s_{n}^m$ BSC. Baresch \textit{et al.} solved the scattering problem for an elastic sphere insonified by an arbitrary incident beam and showed that the problem degenerates to the one of the scattering of an incident plane wave, so that the  corresponding partial wave coefficients $A_{n}$ can be used leading to the relation: $s_{n}^m = A_{n} a_{n}^m$.
(ii) Silva uses the orthogonality and recursion relationship of normalized spherical harmonics directly based on Arfken's textbook (see Appendix \ref{Appendix A}) \cite{arfken2013mathematical}, while Baresch \textit{et al.} use the orthogonality relationship of associated Legendre functions ($P_n^m$) and exponential functions, and also the recursion relationship of associated Legendre functions.
This leads to the fact that there are four terms for the lateral forces and two terms for the axial force in Silva's work (without reindexing) \cite{silva2011expression}, while only two terms for the lateral and one term for the axial forces by Zhao \textit{et al.} \cite{zhao2019computation} (with reindexing during the derivation procedure) \cite{baresch2013three}.
(iii) Silva uses the normalized spherical harmonics, while Baresch \textit{et al.} use the unnormalized spherical harmonics to derive the ARF formulas \cite{baresch2013three}, which have been re-organized with normalized spherical harmonics to be compact by Zhao, Thomas \& Marchiano [Eqs. (1-3)] \cite{zhao2019computation}.

\subsubsection{\label{ARF Gong} Compact expression of the ARF for arbitrary shaped particles.}

If we substitute the relation $s_{n}^m = A_{n}^m a_{n}^m$ for a particle with an arbitrary shape, the correct version of ARF formulas in terms of $a_n^m$ and $s_n^m$ in Eq. (\ref{ARF gong baudoin}) is further written in a compact manner as:
\begin{subequations}
\begin{eqnarray}
F_{x}&=&\frac{\rho_{0} \Phi_{0}^{2}}{4}  \operatorname{Im} \left\{ \sum_{n=0}^{\infty} \sum_{m=-n}^{n}  \left(C_n^{m-1} b_{n+1}^{-m} a_{n}^m a_{n+1}^{m-1*} - C_n^{m+1} b_{n+1}^{m} a_{n}^m a_{n+1}^{ m+1*} \right) \right\}
, \label{Fx eqa final}
\\
F_{y}&=&\frac{\rho_{0} \Phi_{0}^{2}}{4}  \operatorname{Re} \left\{ \sum_{n=0}^{\infty} \sum_{m=-n}^{n}  \left(C_n^{m-1} b_{n+1}^{-m} a_{n}^m a_{n+1}^{m-1*}  + C_n^{m+1} b_{n+1}^{m} a_{n}^m a_{n+1}^{m+1*} \right) \right\}
, \label{Fy eqb final}
\\
F_{z}&=&\frac{\rho_{0} \Phi_{0}^{2}}{2}  \operatorname{Im} \left\{ \sum_{n=0}^{\infty} \sum_{m=-n}^{n} C_n^{m} c_{n+1}^{m} a_{n}^m a_{n+1}^{m*} \right\}
. \label{Fz eqc final}
\end{eqnarray}
\label{ARF anm Anm}
\end{subequations}
with $C_{n}^{m \mp 1} = A_{n}^m+2 A_{n}^m A_{n+1}^{m \mp 1 *}+A_{n+1}^{m \mp 1 *}$ and $C_{n}^m = A_{n}^m+2 A_{n}^m A_{n+1}^{m*}+A_{n+1}^{m*}$. 
Note that these compact equations (\ref{ARF anm Anm}) are equivalent to the re-organized ones (using normalized spherical harmonics instead of spherical harmonics in Ref. \cite{baresch2013three}) by Zhao \textit{et al.} in a direct way \cite{zhao2019computation} when the particle shape is considered as a sphere (so that $A_n^m = A_n$ and $C_n^{m \mp 1} = C_n^m =C_n = A_{n}+2 A_{n} A_{n+1}^*+A_{n+1}^*$) and the index $m$ is with the right regime $|m| \leq n$.

%-----------------------------------------------------------------
\subsection{\label{analyssi of ARF equivalence} Equivalence analysis of the three sets of ARF formulas}

As claimed above, the different forms of ARF formulas derived by Thomas and colleagues \cite{baresch2013three,zhao2019computation} (compact form of correct version of ARF formulas by Silva \cite{silva2011expression} and Gong \textit{et al.} \cite{gong2019t}) and Sapozhnikov \& Bailey \cite{sapozhnikov2013radiation} come from the different elementary wave expansion of velocity potential or pressure. The explicit relation between the beam coefficient $a_n^m$ based on MEM and $H_{nm}$ based on ASM is given by Eq. (\ref{a_nm & H_nm}) in Sec. \ref{sec:relation between Hnm and Anm}, which can be used to substitute into Eq. (\ref{ARF anm Anm}) to derive the 3D ARF formulas in terms of the notation $H_{nm}$ introduced by Sapozhnikov \& Bailey based on the ASM.
The equivalence between the two sets of formulas will be verified immediately if $A_n^m =A_n$ is set for a spherical shape (see details in Appendix \ref{Appendix C}).
The question raised by Sapozhnikov \& Bailey in their paper \cite{sapozhnikov2013radiation} between their formula and the one by Silva is now solved. All in all, considering the correction of the index issues pointed out above, all the three sets of original 3D ARF formulas are proved to be equivalent.

%-----------------------------------------------------------------
\section{\label{sec:equivalence ART}Equivalence of three-dimensional ART formulas}

The ART on a particle in an ideal fluid can be calculated by the integral of the time-averaged angular stress tensor minus the angular momentum flux over a far-field standard spherical shape centered at the mass center of the particle \cite{maidanik1958torques,zhang2011angular,silva2012radiation,gong2019reversals} (see Eq. \ref{torque in potential}). Explicit expressions of 3D ART formulas have been derived by Silva \textit{et al.} \cite{silva2012radiation} and Gong \textit{et al.} \cite{gong2019reversals} based on the MEM and Gong \& Baudoin based on the ASM \cite{gong2020ART}.

 As for the ARF, there are also index issues in the expression obtained by Silva \textit{et al.} \cite{silva2012radiation} and Gong \textit{et al.} \cite{gong2019reversals}. Here we provide the correct expressions of ART formulas by Silva \textit{et al.} \cite{silva2012radiation} and Gong \textit{et al.} \cite{gong2019reversals} based on the multipole expansion method (see details in Appendix \ref{Appendix E}):
\begin{subequations}
\begin{eqnarray}
T_{x}&=&-\frac{\rho_{0} \Phi_{0}^{2}}{4 k} \operatorname{Re}\left\{\sum_{n=0}^{\infty} \sum_{m=-n+1}^{n} \overline{b}_{n}^m \left[\left(a_{n}^{m*}+s_{n}^{m*}\right) s_{n}^{m-1}+\left(a_{n}^{m-1*}+s_{n}^{m-1*}\right) s_{n}^m\right]\right\}, \label{MEM_x anm snm}
\\
T_{y}&=&-\frac{\rho_{0} \Phi_{0}^{2}}{4 k} \operatorname{Im}\left\{\sum_{n=0}^{\infty} \sum_{m=-n+1}^{n} \overline{b}_{n}^m \left[\left(a_{n}^{m*}+s_{n}^{m*}\right) s_{n}^{m-1}-\left(a_{n}^{m-1*}+s_{n}^{m-1*}\right) s_{n}^ m\right]\right\}, 
\label{MEM_y anm snm}
\\
T_{z}&=&-\frac{\rho_{0} \Phi_{0}^{2}}{2 k} \operatorname{Re} \left\{ \sum_{n=0}^{\infty} \sum_{m=-n}^{n} m\left(a_{n}^{m*}+s_{n}^{m*}\right) s_{n}^m \right\}, \label{MEM_z}
\end{eqnarray}
\label{ART anm snm}
\end{subequations}
with the coefficients $\overline{b}_{n}^m$ given in Appendix \ref{Appendix D}. Again, it is noteworthy that the numerical computations in Ref. \cite{gong2019reversals} are correct since they use the definition for the scattered BSC that $s_n^m = 0$ when $n<0$ or $|m|>n$.

The relationship $s_n^m = A_n^m a_n^m$ can be introduced into Eq. (\ref{ART anm snm}) to obtain a set of compact formulas in terms of the incident BSC only:

\begin{subequations}
\begin{eqnarray}
T_{x}&=&-\frac{\rho_{0} \Phi_{0}^{2}}{4 k} \operatorname{Re}\left\{\sum_{n=0}^{\infty} \sum_{m=-n+1}^{n}
\overline{b}_{n}^m \overline{C}_{n}^m a_{n}^{m*} a_{n}^{m-1}\right\}, \label{MEM_x anm}
\\
T_{y}&=&-\frac{\rho_{0} \Phi_{0}^{2}}{4 k} \operatorname{Im} \left\{\sum_{n=0}^{\infty} \sum_{m=-n+1}^{n} \overline{b}_{n}^m \overline{C}_{n}^m a_{n}^{ m*} a_{n}^{m-1}\right\}, \label{MEM_y anm}
\\
T_{z}&=&-\frac{\rho_{0} \Phi_{0}^{2}}{2 k}\operatorname{Re} \left\{\sum_{n=0}^{\infty} \sum_{m=-n}^{n} m \overline{D}_{n}^m a_{n }^{m*} a_{n}^m \right\}. \label{MEM_z anm}
\end{eqnarray}
\label{MEM Torque}
\end{subequations}
where $\overline{C}_{n}^m = A_{n}^{m-1}+A_{n}^{m*}+2 A_{n}^{m-1} A_{n}^{m*}$, $\overline{D}_{n}^m = A_{n}^m + A_{n}^m A_{n}^{m*}$. The above compact ART formulas are identical to Eqs. (10-12) of Ref. \cite{gong2020ART} by using the relation between $a_n^m$ and $H_{nm}$ given by Eq. (\ref{a_nm & H_nm}) in Sec \ref{sec:relation between Hnm and Anm} (see details in Appendix \ref{Appendix F}). Hence, the equivalence of the ART formulas between the correct form [see Eq. (\ref{ART anm snm})] of Silva \textit{et al.} and Gong \textit{et al.}'s work based on the MEM \cite{silva2012radiation,gong2019reversals} and those derived by Gong and Baudoin based on the ASM \cite{gong2020ART} is demonstrated in this section.

%---------------------------------------------------------------------
\section{\label{sec: conclusion} Conclusions and discussions}

In summary, we provide in this paper a clear proof of the equivalence of the three sets of the 3D acoustic radiation force (ARF) formulas derived independently by Silva \cite{silva2011expression} (extended later on by Gong \textit{et al.} \cite{gong2019t} to arbitrary shape particles), Thomas and associates \cite{baresch2013three,zhao2019computation}, 
and Sapozhnikov \& Bailey \cite{sapozhnikov2013radiation}, and the 3D acoustic radiation torque (ART) formulas derived by Silva \textit{et al.} \cite{silva2012radiation} (extended by Gong \textit{et al.} \cite{gong2019reversals} to arbitrary shape particles) and Gong \& Baudoin \cite{gong2020ART}. The reasons for the different forms of ARF and ART expressions are discussed completely in Secs. \ref{arf equivalence} and \ref{sec:equivalence ART}, respectively.

The advantage of the MEM-based ARF and ART formulas \cite{silva2011expression,baresch2013three,gong2019t} is that the calculations of 3D ARF and ART are direct by using the incident BSC $a_n^m$ of a known acoustic field on a particle with available $A_n^m$ which has a long research history in the literature for scattering problems. For a non planar beam such as Bessel beam, the incident BSC $a_n^m$ is affected by the relative position of the beam axis with respect to the particle center. In Silva's work, the off-axis incidence of a Bessel beam was not studied  \cite{silva2011expression}. This was accomplished by Baresch \textit{et al.} \cite{baresch2013three} and Gong \textit{et al.} \cite{gong2019t} with two different ideas: Under the off-axis incidence, there are two way to calculate the incident BSC $a_n^m$: (i) the first one is to involve the off-set information inside the elementary wave [i.e., $j_n(kr)Y_n^m(\theta,\phi)$] as demonstrated by Baresch \textit{et al.} \cite{baresch2013three} using the translocation and rotation matrices for spherical harmonics and illustrated on the example of a helicoidal Bessel beam; (ii) the second one is to put the off-set information in the BSC $a_n^m$ directly using the addition theorem for the Bessel functions for a cylindrical Bessel beam by Gong \textit{et al.} \cite{gong2017multipole}, which is limited to certain kinds of ideal beams. The latter way can be taken as a special case of the former. The $a_n^m$ can be also calculated through numerical integration \cite{silva2011off}, however, with the drawback of the computational cost and parameter selection of the beams \cite{gong2019resonance}.

The advantage of the ASM-based ARF \cite{sapozhnikov2013radiation} and ART \cite{gong2020ART} is that they are easy to set up when the field is known (e.g. measured) in a transverse plane (e.g. for planar holographic transducers \cite{pre_jimenez_2016,prap_riaud_2017,apl_jimenez_2018,sa_baudoin_2019,baudoin2020naturecell}).  Note that this set of formulas can also be used for ideal beams whose introduced coefficients $H_{nm}$ are available either by using the angular spectrum of the beam $S({k_x,k_y)}$ \cite{sapozhnikov2013radiation} or the relation given in Eq. (\ref{a_nm & H_nm}).

To finalize the calculation of the ARF and ART with all these formulas, the key point is to obtain the partial wave coefficients $A_n^m$ of the particle exactly. Silva \cite{silva2011expression,silva2012radiation}, Thomas and colleagues \cite{baresch2013three,zhao2019computation} and Sapozhnikov \& Bailey \cite{sapozhnikov2013radiation} discuss particles with spherical shapes so that $A_n^m$ only depends on the index $n$, having $A_n^m = A_n$. 
Gong \textit{et al.} derive the formulas with $A_n^m$ depending on the indexes of $(n,m)$ with several numerical computation for arbitrary-sized nonspherical shapes by a semi-analytical T-matrix method \cite{gong2019t,gong2019reversals,gong2018thesis}. For a rigid spheroidal particle in the so-called long-wavelength limit, Silva and colleagues gives the $A_n^m$ with the Taylor expansion up to the dipole ($n=1$) in spheroidal coordinates \cite{silva2018acoustic} and obtain concise analytical ARF  and ART expressions using the partial wave expansion  \cite{silva2020nonlinear}.  Note also that the overall formulas discussed in the present work are generally applied for a particle in an ideal fluid but are still applicable for a particle in a viscous fluid if the viscous effect in the fluid can be accounted in the expression of scattering (partial wave) coefficients
\cite{zhang2011angular,baresch2018orbital}.

From a perspective viewpoint, the present work on the ARF and ART formulas may be extended for multiple particles \cite{bostrom1980multiple,silva2014acoustic} if the partial wave coefficients are available, which can be used for the manipulation and assembly of large particles beyond Rayleigh regime \cite{gong2019particle,prap_gong_2020}. 
Based on the Eqs. (\ref{ARF anm Anm}) and (\ref{MEM Torque}), the ARF and ART are closely related to the scattering from the particle in a fluid. 
Hence, the scattering characteristics are essential to the acting force and torque of acoustic field on the particle. 
For example, the resonance scattering from an elastic sphere may be suppressed under a on-axis Bessel beam of selected parameters and be not with an off-axis incidence \cite{marston2007acoustic,marston2009erratum,gong2019resonance}, which could be used to tune the ARF and ART, such as a stable tractor (pulling) beam \cite{marston2006axial,zhang2011geometrical,fan2019trapping}, or a 3D stable trapping \cite{baresch2013spherical,baresch2016observation} with suppressed spinning rotation.
More importantly, the present work will help to build acoustical tweezers numerical toolbox \cite{gong2018conference,gong2018thesis} as an analogy to its optical counterpart \cite{nieminen2007optical}.

%---------------------------------------------------------------------
\appendix
\section{\label{Appendix A} orthogonality and recurrence relations of spherical harmonics}
The orthogonality relationship of normalized spherical harmonics is given in Eq. (15.138) by Arfken \textit{\textit{et al.}} \cite{arfken2013mathematical}

\begin{equation}
\int_{0}^{2 \pi} d \varphi \int_{0}^{\pi} \sin \theta d \theta Y_{n}^{m*} Y_{n'}^{m'} = \delta_{nn'} \delta_{mm'},
\label{Ynm Ynm}
\end{equation}

The recurrence relations of normalized spherical harmonics involved with trigonometric and exponential functions are given in Eqs. (15.150) and (15.151) by Arfken \textit{\textit{et al.}} \cite{arfken2013mathematical} , respectively

\begin{equation}
\cos \theta Y_{n}^m = c_{n}^{m} Y_{n-1}^{m}+c_{n+1}^{m} Y_{n+1}^{m},
\label{cos Ynm}
\end{equation}

with 
\begin{equation}
c_{n}^m = \sqrt{\frac{(n+m)(n-m)}{(2 n-1)(2 n+1)}},
\label{cnm}
\end{equation}
which is based on a recurrence relation of associated Lengendre functions [Eq. (15.88) in Arfken \textit{\textit{et al.}}'s textbook], as also used by Baresch \textit{et al.} in Eq. (C5) in Appendix C \cite{baresch2013three}. And
\begin{equation}
e^{\pm i \varphi} \sin \theta Y_{n}^m = \pm b_{n}^{\mp m-1} Y_{n-1}^{m \pm 1} \mp b_{n+1}^{\pm m} Y_{n+1}^{m \pm 1}
\label{exp Ynm}
\end{equation}
with
\begin{equation}
b_{n}^m = \sqrt{\frac{(n+m)(n+m+1)}{(2 n-1)(2 n+1)}},
\label{bnm}
\end{equation}
which is based on two recurrence relations of associated Lengendre functions [Eqs. (15.89-90) in Arfken \textit{\textit{et al.}}'s textbook], with Eq. (15.89) also used by Baresch \textit{et al.} in Appendix D \cite{baresch2013three}.

By using the Euler's formula $e^{\pm i \varphi}=\cos \varphi \pm i \sin \varphi$, the terms of normalized spherical harmonics involved with trigonometric functions ($\cos \varphi \sin \theta Y_{nm}$ and $\sin \varphi \sin \theta Y_{nm}$) can be obtained, which can be further applied into Eq. (11) in Gong \textit{\textit{et al.}} \cite{gong2019t} for the final 3D ARF expressions. The relation used for the derivation of $F_x$
\begin{equation}
2 \cos \varphi \sin \theta Y_{n}^{m} =
b_{n}^{-m-1} Y_{n-1}^{m+1} - b_{n+1}^{m} Y_{n+1}^{m+1}+b_{n+1}^{-m} Y_{n+1}^{m-1}-b_{n}^{m-1} Y_{n-1}^{m-1}.
\label{cos phi Ynm}
\end{equation}
and the expression for the derivation of $F_y$
\begin{equation}
2 i \times \sin \varphi \sin \theta Y_{n}^{m} = b_{n}^{-m-1} Y_{n-1}^{m+1}-b_{n+1}^{m} Y_{n+1}^{m+1}-b_{n+1}^{-m} Y_{n+1}^{m-1}+b_{n}^{m-1} Y_{n-1}^{m-1}.
\label{isin phi Ynm}
\end{equation}

%---------------------------------------------------------------------
\section{\label{Appendix B}  Detailed derivation of ARF with correct index}

\subsection{\label{Appendix B1}  Detailed derivation of $F_x$}

Based on the ARF formulas of Eq. (9) from Ref.\cite{gong2019t}, the expression of $x$-component of ARF is
\begin{eqnarray}
\begin{aligned}
F_{x} &=\frac{1}{2} \rho_{0} k^{2} \Phi_{0}^{2} \iint_{S_{0}} \operatorname{Re}\left\{-\sum_{n=0}^{\infty} \sum_{m=-n}^{n} \sum_{n'=0}^{\infty} \sum_{m'=-n'}^{n'} \frac{i^{n'-n}}{(k r)^{2}}\left(a_{n}^m+s_{n}^m \right) s_{n'}^{m'*} Y_{n}^m Y_{n'}^{m'*}\right\} \\
& \times r^{2} \sin \theta \cos \varphi \sin \theta d \theta d \varphi
\\
&=-\frac{1}{2} \rho_{0} \Phi_{0}^{2} \operatorname{Re}\left\{\sum_{n=0}^{\infty} \sum_{m=-n}^{n} \sum_{n'=0}^{\infty} \sum_{m'=-n'}^{n'} i^{n'-n}\left(a_{n}^m+s_{n}^m\right) s_{n'}^{m'*} \iint_{S_{0}} Y_{n}^m Y_{n'}^{m'*} \sin \theta \cos \varphi \sin \theta d \theta d \varphi\right\}
\end{aligned}, \label{Fx in a_nm and s_nm1}
\end{eqnarray}
Substituting  Eqs. (\ref{cos phi Ynm}) into (\ref{Fx in a_nm and s_nm1}), $F_x$ can be divided into 4 terms:
\begin{eqnarray}
\begin{aligned}
F_{x} =& -\frac{1}{4} \rho_{0} \Phi_{0}^{2} \operatorname{Re}\left\{\sum_{n=1}^{\infty} \sum_{m=-n}^{n-2} \sum_{n'=0}^{\infty} \sum_{m'=-n'}^{n'} i^{n'-n}\left(a_{n}^m+s_{n}^m\right) s_{n'}^{m'*} \iint_{S_{0}} b_{n}^{-m-1} Y_{n-1}^{m+1} Y_{n'}^{m'*} \sin \theta d \theta d \varphi \right.
\\
&+\sum_{n=0}^{\infty} \sum_{m=-n}^{n} \sum_{n'=0}^{\infty} \sum_{m'=-n'}^{n'} i^{n'-n}\left(a_{n}^m+s_{n}^m\right) s_{n'}^{m'*} \iint_{S_{0}} -b_{n+1}^{m} Y_{n+1}^{m+1} Y_{n'}^{m'*} \sin \theta d \theta d \varphi
\\
&+\sum_{n=0}^{\infty} \sum_{m=-n}^{n} \sum_{n'=0}^{\infty} \sum_{m'=-n'}^{n'} i^{n'-n}\left(a_{n}^m+s_{n}^m\right) s_{n'}^{m'*} \iint_{S_{0}} b_{n+1}^{-m} Y_{n+1}^{m-1} Y_{n'}^{m'*} \sin \theta d \theta d \varphi
\\
&\left. +\sum_{n=1}^{\infty} \sum_{m=-n+2}^{n} \sum_{n'=0}^{\infty} \sum_{m'=-n'}^{n'} i^{n'-n}\left(a_{n}^m+s_{n}^m\right) s_{n'}^{m'*} \iint_{S_{0}} -b_{n}^{m-1} Y_{n-1}^{m-1} Y_{n'}^{m'*} \sin \theta d \theta d \varphi \right\}
\end{aligned}, \label{Fx in a_nm and s_nm2}
\end{eqnarray}
It is important to note that the index regimes of $(n,m)$ for different terms are different because the correct index regime of $Y_n^m$ should be $n \in [0,\infty]$ and $ m \in [-n, n]$ for the indexes.
In addition, based on the definition in Eq. (\ref{incident potential MEM}), the intersection of regime of $(n,m)$ is listed in Table \ref{table1}. 

\begin{table}[!htbp]
\small
  \caption{Regime of $(n,m)$  in normalized spherical harmonics for derivaiton of $F_x$ and $F_y$. Note that based on the definition in Eq. (\ref{potential MEM}), we have $n \in [0,\infty]$ and $m \in [-n,n]$.}
  \label{table1}
  \begin{tabular}{c |c | c | c}
\hline
 &  $n$ &  $m$ & Intersection \\
\hline 
$Y_{n-1}^{m+1}$ & $n \in [1,\infty]$ & $ m \in [-n, n-2]$ & $n \in [1,\infty]$, $ m \in [-n, n-2]$\\

$Y_{n+1}^{m+1}$ & $n \in [-1,\infty]$ & $ m \in [-n-2, n]$ & $n \in [0,\infty]$, $ m \in [-n, n]$\\

$Y_{n+1}^{m-1}$ & $n \in [-1,\infty]$ & $ m \in [-n, n+2]$ & $n \in [0,\infty]$, $ m \in [-n, n]$\\

$Y_{n-1}^{m-1}$ & $n \in [1,\infty]$ & $ m \in [-n+2, n]$ & $n \in [1,\infty]$, $ m \in [-n+2, n]$\\
 \hline
  \end{tabular}
\end{table}

Using the orthogonality relation of the normalized spherical harmonics in Eq. (\ref{Ynm Ynm}), Eq.(\ref{Fx in a_nm and s_nm2}) can be further written as
\begin{eqnarray}
\begin{aligned}
F_{x} =& -\frac{1}{4} \rho_{0} \Phi_{0}^{2} \operatorname{Re}\left\{
\sum_{n=1}^{\infty} \sum_{m=-n}^{n-2} \sum_{n'=0}^{\infty} \sum_{m'=-n'}^{n'} i^{n'-n}\left(a_{n}^m+s_{n}^m\right) s_{n'}^{m'*} b_{n}^{-m-1} \delta_{n-1,n'} \delta_{m+1,m'}  \right.
\\
&+\sum_{n=0}^{\infty} \sum_{m=-n}^{n} \sum_{n'=0}^{\infty} \sum_{m'=-n'}^{n'} - i^{n'-n} \left(a_{n}^m+s_{n}^m\right) s_{n'}^{m'*} b_{n+1}^{m} \delta_{n+1,n'} \delta_{m+1,m'}
\\
&+\sum_{n=0}^{\infty} \sum_{m=-n}^{n} \sum_{n'=0}^{\infty} \sum_{m'=-n'}^{n'} i^{n'-n}\left(a_{n}^m+s_{n}^m\right) s_{n'}^{m'*} b_{n+1}^{-m} \delta_{n+1,n'} \delta_{m-1,m'} 
\\
& \left. +\sum_{n=1}^{\infty} \sum_{m=-n+2}^{n} \sum_{n'=0}^{\infty} \sum_{m'=-n'}^{n'} -i^{n'-n}\left(a_{n}^m+s_{n}^m\right) s_{n'}^{m'*} b_{n}^{m-1} \delta_{n-1,n'} \delta_{m-1,m'} \right\}
\\
=&  -\frac{1}{4} \rho_{0} \Phi_{0}^{2} \operatorname{Re}\left\{\sum_{n=1}^{\infty} \sum_{m=-n}^{n-2} -i\left(a_{n}^m+s_{n}^m\right) s_{n-1}^{m+1*} b_{n}^{-m-1} +\sum_{n=0}^{\infty} \sum_{m=-n}^{n} 
- i \left(a_{n}^m+s_{n}^m\right) s_{n+1}^{m+1*} b_{n+1}^{m} \right.\\
& \left. +\sum_{n=0}^{\infty} \sum_{m=-n}^{n}  i\left(a_{n}^m+s_{n}^m\right) s_{n+1}^{m-1*} b_{n+1}^{-m}+\sum_{n=1}^{\infty} \sum_{m=-n+2}^{n} i\left(a_{n}^m+s_{n}^m\right) s_{n-1}^{m-1*} b_{n}^{m-1}
\right\}
\\
=&  -\frac{1}{4} \rho_{0} \Phi_{0}^{2} \operatorname{Im}\left\{\sum_{n=1}^{\infty} \sum_{m=-n}^{n-2} \left(a_{n}^m+s_{n}^m\right) s_{n-1}^{m+1*} b_{n}^{-m-1} +\sum_{n=0}^{\infty} \sum_{m=-n}^{n} 
 \left(a_{n}^m+s_{n}^m\right) s_{n+1}^{m+1*} b_{n+1}^{m} \right.\\
& \left. +\sum_{n=0}^{\infty} \sum_{m=-n}^{n}  -\left(a_{n}^m+s_{n}^m\right) s_{n+1}^{m-1*} b_{n+1}^{-m}+\sum_{n=1}^{\infty} \sum_{m=-n+2}^{n} -\left(a_{n}^m+s_{n}^m\right) s_{n-1}^{m-1*} b_{n}^{m-1}
\right\}
\end{aligned} \label{Fx in a_nm and s_nm3}
\end{eqnarray}
Note that Re$\{X\}=$ Im$\{iX\}$ with $X$ an arbitrary complex number.
Here, a re-index  is applied with $p=n-1 \in [0,\infty]$ for the first and fourth term of Eq. (\ref{Fx in a_nm and s_nm3})
\begin{eqnarray}
\begin{aligned}
F_{x} =&  -\frac{1}{4} \rho_{0} \Phi_{0}^{2} \operatorname{Im}\left\{\sum_{p=0}^{\infty} \sum_{m=-p-1}^{p-1} \left(a_{p+1}^m+s_{p+1}^m\right) s_{p}^{m+1*} b_{p+1}^{-m-1} +\sum_{n=0}^{\infty} \sum_{m=-n}^{n} 
 \left(a_{n}^m+s_{n}^m\right) s_{n+1}^{m+1*} b_{n+1}^{m} \right.\\
& \left. +\sum_{n=0}^{\infty} \sum_{m=-n}^{n}  -\left(a_{n}^m+s_{n}^m\right) s_{n+1}^{m-1*} b_{n+1}^{-m}+\sum_{p=0}^{\infty} \sum_{m=-p+1}^{p+1} -\left(a_{p+1}^m+s_{p+1}^m\right) s_{p}^{m-1*} b_{p+1}^{m-1}
\right\}
\end{aligned} \label{Fx in a_nm and s_nm4}
\end{eqnarray}

Now, using a re-index for $m$: for the first term $q=m+1 \in [-p,p]$, and for the fourth term  $q=m-1 \in [-p,p]$, we have 

\begin{eqnarray}
\begin{aligned}
F_{x} =&  -\frac{1}{4} \rho_{0} \Phi_{0}^{2} \operatorname{Im}\left\{\sum_{p=0}^{\infty} \sum_{q=-p}^{p} \left(a_{p+1}^{q-1}+s_{p+1}^{q-1}\right) s_{p}^{q*} b_{p+1}^{-q} +\sum_{n=0}^{\infty} \sum_{m=-n}^{n} 
 \left(a_{n}^m+s_{n}^m\right) s_{n+1}^{m+1*} b_{n+1}^{m} \right.\\
& \left. +\sum_{n=0}^{\infty} \sum_{m=-n}^{n}  -\left(a_{n}^m+s_{n}^m\right) s_{n+1}^{m-1*} b_{n+1}^{-m}+\sum_{p=0}^{\infty} \sum_{q=-p}^{p} -\left(a_{p+1}^{q+1}+s_{p+1}^{q+1}\right) s_{p}^{q*} b_{p+1}^{q} 
\right\}
\\
=&  \frac{1}{4} \rho_{0} \Phi_{0}^{2} \operatorname{Im}\left\{\sum_{n=0}^{\infty} \sum_{m=-n}^{n} \Big[ b_{n+1}^{-m} \big[ \left(a_{n}^m+s_{n}^m\right) s_{n+1}^{m-1*}  - \left(a_{n+1}^{m-1}+s_{n+1}^{m-1}\right) s_{n}^{m*} \big]  \right.\\
& \left. + b_{n+1}^{m} \big[ \left(a_{n+1}^{m+1}+s_{n+1}^{m+1}\right) s_{n}^{m*} -  \left(a_{n}^m+s_{n}^m\right) s_{n+1}^{m+1*} \big]  \Big] \right.\Bigg\}
\end{aligned} \label{Fx in a_nm and s_nm5}
\end{eqnarray}
which is Eq. (\ref{Fx gong baudoin a}) in Sec. \ref{Silva's ARF}.

\subsection{\label{Appendix B2}  Derivation of $F_y$}

The expression of $y$-component of ARF is
\begin{eqnarray}
\begin{aligned}
F_{y} &=\frac{1}{2} \rho_{0} k^{2} \Phi_{0}^{2} \iint_{S_{0}} \operatorname{Re}\left\{-\sum_{n=0}^{\infty} \sum_{m=-n}^{n} \sum_{n'=0}^{\infty} \sum_{m'=-n'}^{n'} \frac{i^{n'-n}}{(k r)^{2}}\left(a_{n}^m+s_{n}^m\right) s_{n' }^{m'*} Y_{n }^m(\theta, \varphi) Y_{n' }^{m'*}(\theta, \varphi)\right\} \\
& \times r^{2} \sin \theta \sin \varphi \sin \theta d \theta d \varphi
\\
&=-\frac{1}{2} \rho_{0} \Phi_{0}^{2} \operatorname{Re}\left\{\sum_{n=0}^{\infty} \sum_{m=-n}^{n} \sum_{n'=0}^{\infty} \sum_{m'=-n'}^{n'} i^{n'-n}\left(a_{n}^m+s_{n}^m\right) s_{n'}^{m'*} \iint_{S_{0}} Y_{n}^m Y_{n'}^{m'*} \sin \theta \sin \varphi \sin \theta d \theta d \varphi\right\}
\end{aligned}, \label{Fy in a_nm and s_nm}
\end{eqnarray}
The detailed derivation of $F_y$ is similar as that for the $x$-component $F_x$ by substituting Eq. (\ref{isin phi Ynm}) replacing of (\ref{cos phi Ynm}) into (\ref{Fy in a_nm and s_nm}), and using of the orthogonality relationship of normalized spherical harmonics of Eq. (\ref{Ynm Ynm}). The final expression of $F_y$ in terms of $a_n^m$ and $s_n^m$ is given in Eq. (\ref{Fy gong baudoin b}) in Sec. \ref{Silva's ARF}, which is not given here for brevity.

\subsection{\label{Appendix B3}  Detailed derivation of $F_z$}

The expression of $z$-component of ARF is
\begin{eqnarray}
\begin{aligned}
F_{z} =& \frac{1}{2} \rho_{0} k^{2} \Phi_{0}^{2} \iint_{S_{0}} \operatorname{Re}\left\{-\sum_{n=0}^{\infty} \sum_{m=-n}^{n} \sum_{n'=0}^{\infty} \sum_{m'=-n'}^{n'} \frac{i^{n'-n}}{(k r)^{2}}\left(a_{n}^m+s_{n}^m\right) s_{n' }^{m'*} Y_{n }^m(\theta, \varphi) Y_{n' }^{m'*}(\theta, \varphi)\right\} \\
& \times r^{2} \cos \theta \sin \theta d \theta d \varphi
\\
=&-\frac{1}{2} \rho_{0} \Phi_{0}^{2} \operatorname{Re}\left\{\sum_{n=0}^{\infty} \sum_{m=-n}^{n} \sum_{n'=0}^{\infty} \sum_{m'=-n'}^{n'} i^{n'-n}\left(a_{n}^m+s_{n}^m\right) s_{n'}^{m'*} \iint_{S_{0}} Y_{n}^m Y_{n'}^{m'*} \cos \theta \sin \theta d \theta d \varphi\right\}
\end{aligned}, \label{Fz in a_nm and s_nm1}
\end{eqnarray}
Substituting Eq. (\ref{cos Ynm}), we have
\begin{eqnarray}
\begin{aligned}
F_{z} =&-\frac{1}{2} \rho_{0} \Phi_{0}^{2} \operatorname{Re}\left\{\sum_{n=1}^{\infty} \sum_{m=-n+1}^{n-1} \sum_{n'=0}^{\infty} \sum_{m'=-n'}^{n'} i^{n'-n}\left(a_{n}^m+s_{n}^m\right) s_{n'}^{m'*} \iint_{S_{0}} c_{n}^{m} Y_{n-1}^{m} Y_{n'}^{m'*} \sin \theta d \theta d \varphi \right.\\
& \left. +\sum_{n=0}^{\infty} \sum_{m=-n}^{n} \sum_{n'=0}^{\infty} \sum_{m'=-n'}^{n'} i^{n'-n}\left(a_{n}^m+s_{n}^m\right) s_{n'}^{m'*} \iint_{S_{0}} c_{n+1}^{m} Y_{n+1}^{m} Y_{n'}^{m'*} \sin \theta d \theta d \varphi
\right\}
\end{aligned}, \label{Fz in a_nm and s_nm2}
\end{eqnarray}
For the definition of $(n,m)$ in Eq. (\ref{incident potential MEM}), it has $n \in [0,\infty]$ and $m \in [-n,n]$. Since $Y_{n-1}^m$ ($n \in  [1,\infty]$ and $m \in [-n+1,n-1]$) and $Y_{n+1}^m$ ($n \in [-1,\infty]$ and $m \in [-n-1,n+1]$) are introduced here, the final regimes of indexes $(n,m)$ are the intersection and given differently for the first and second part.

\begin{table}[!htbp]
\small
  \caption{Regime of $(n,m)$  in normalized spherical harmonics for derivation of $F_z$. Note that based on the definition in Eq. (\ref{incident potential MEM}), we have $n \in [0,\infty]$ and $m \in [-n,n]$.}
  \label{table2}
  \begin{tabular}{c |c | c | c}
\hline
  &  $n$ &  $m$ & Intersection \\
\hline 
$Y_{n-1}^{m+1}$ & $n \in [1,\infty]$ & $ m \in [-n, n-2]$ & $n \in [1,\infty]$, $ m \in [-n, n-2]$\\

$Y_{n+1}^{m+1}$ & $n \in [-1,\infty]$ & $ m \in [-n-2, n]$ & $n \in [0,\infty]$, $ m \in [-n, n]$\\

 \hline
  \end{tabular}
\end{table}

By using Eq. (\ref{Ynm Ynm}), the expression of $F_z$ is
\begin{eqnarray}
\begin{aligned}
F_{z} =&-\frac{1}{2} \rho_{0} \Phi_{0}^{2} \operatorname{Re}\left\{\sum_{n=1}^{\infty} \sum_{m=-n+1}^{n-1} \sum_{n'=0}^{\infty} \sum_{m'=-n'}^{n'} i^{n'-n}\left(a_{n}^m+s_{n}^m\right) s_{n'}^{m'*} c_{n}^{m} \delta_{n-1, n'} \delta_{m, m'} \right.\\
& \left. +\sum_{n=0}^{\infty} \sum_{m=-n}^{n} \sum_{n'=0}^{\infty} \sum_{m'=-n'}^{n'} i^{n'-n}\left(a_{n}^m+s_{n}^m\right) s_{n'}^{m'*} c_{n+1}^{m} \delta_{n+1, n'} \delta_{m, m'}
\right\}
\\
=&-\frac{1}{2} \rho_{0} \Phi_{0}^{2} \operatorname{Re}\left\{\sum_{n=1}^{\infty} \sum_{m=-n+1}^{n-1}  i^{-1}\left(a_{n}^m+s_{n}^m\right) s_{n-1}^{m*} c_{n}^{m} 
+ \sum_{n=0}^{\infty} \sum_{m=-n}^{n}  i\left(a_{n}^m+s_{n}^m\right) s_{n+1}^{m*} c_{n+1}^{m} 
\right\}
\end{aligned}, \label{Fz in a_nm and s_nm3}
\end{eqnarray}

Re-index for the first part of Eq. (\ref{Fz in a_nm and s_nm3}) using $p=n-1 \in [0,\infty]$, so that $m  \in [-p,p]$. The final form of $F_z$ in terms of $a_n^m$ and $s_n^m$ is 
\begin{eqnarray}
\begin{aligned}
F_{z} =&-\frac{1}{2} \rho_{0} \Phi_{0}^{2} \operatorname{Re}\left\{\sum_{p=0}^{\infty} \sum_{m=-p}^{p}  i^{-1}\left(a_{p+1}^m+s_{p+1}^m\right) s_{p}^{m*} c_{p+1}^{m} 
+ \sum_{n=0}^{\infty} \sum_{m=-n}^{n}  i\left(a_{n}^m+s_{n}^m\right) s_{n+1}^{m*} c_{n+1}^{m} \right\}
\\
=&-\frac{1}{2} \rho_{0} \Phi_{0}^{2} \operatorname{Re}\left\{\sum_{n=0}^{\infty} \sum_{m=-n}^{n}  i\left(a_{n+1}^{m*}+s_{n+1}^{m*}\right) s_{n}^{m} c_{n+1}^{m} 
+ \sum_{n=0}^{\infty} \sum_{m=-n}^{n}  i\left(a_{n}^m+s_{n}^m\right) s_{n+1}^{m*} c_{n+1}^{m} \right\}
\\
=&\frac{1}{2} \rho_{0} \Phi_{0}^{2} \operatorname{Im}\left\{\sum_{n=0}^{\infty} \sum_{m=-n}^{n} c_{n+1}^{m} \Big[ \left(a_{n+1}^{m*}+s_{n+1}^{m*}\right) s_{n}^{m}  
+  \left(a_{n}^m+s_{n}^m\right) s_{n+1}^{m*} \Big] \right\}
\end{aligned}. \label{Fz in a_nm and s_nm4}
\end{eqnarray}
which is Eq. (\ref{Fz gong baudoin c}) in Sec. \ref{Silva's ARF}. Note that Re$\{X\}$=Re$\{X^*\}$ and Re$\{iX\}$=$-$Im$\{X\}$. 

%---------------------------------------------------------------------
\section{\label{Appendix C}  Equivalence of Eq. (11) and formulas by Sapozhnikov \& Bailey}
By substituting Eq. (\ref{a_nm & H_nm}) into (\ref{ARF anm Anm}), we can prove that the three components of ARF formulas for a sphere (with $A_n^m = A_n$) are equivalent to those in terms of $H_{nm}$ by Sapozhnikov \& Bailey [see Eqs. (46-48) in Ref. \cite{sapozhnikov2013radiation}], respectively. The detailed derivations are given below.

Recall that for a sphere, one has $C_n^{m \mp 1}$ = $C_n$. The $x$-component of ARF: 
\begin{eqnarray}
\begin{aligned}
F_{x}&=\frac{\rho_{0} \Phi_{0}^{2}}{4}  \operatorname{Im} \left\{ \sum_{n=0}^{\infty} \sum_{m=-n}^{n} C_{n} \left( b_{n+1}^{-m} a_{n}^m a_{n+1}^{m-1*} - b_{n+1}^{m} a_{n}^m a_{n+1}^{ m+1*} \right) \right\}
\\
&=\frac{1}{4 \pi^{2} \rho_{0} k^{2} c^{2}}  \operatorname{Im} \left\{ \sum_{n=0}^{\infty} \sum_{m=-n}^{n} C_{n} \left[ b_{n+1}^{-m} 
\big(i^{n-1} H_{n m}\big)
\big(i^{n} H_{n+1, m-1}\big)^* 
- b_{n+1}^{m} \big( i^{n-1} H_{n m}\big) \big( i^{n} H_{n+1, m+1}\big)^*
 \right] \right\}
\\
&=\frac{1}{4 \pi^{2} \rho_{0} k^{2} c^{2}}  \operatorname{Im} \left\{ \sum_{n=0}^{\infty} \sum_{m=-n}^{n} i C_{n} \left(-b_{n+1}^{-m} 
H_{n m} H_{n+1, m-1}^*
+ b_{n+1}^{m} H_{n m} H_{n+1, m+1}^*
 \right) \right\}
\\
&=\frac{1}{4 \pi^{2} \rho_{0} k^{2} c^{2}}  \operatorname{Re} \left\{ \sum_{n=0}^{\infty} \sum_{m=-n}^{n} C_{n} \left(-b_{n+1}^{-m} 
H_{n m} H_{n+1, m-1}^*
+ b_{n+1}^{m} H_{n m} H_{n+1, m+1}^*
 \right) \right\}
\end{aligned}
\label{Fx in H_nm}
\end{eqnarray}
Note that $\omega = kc$ with the sound speed in fluid $c$, and Im$\{iX\}=$ Re$\{X\}$. By replacing $-m$ with $m$ for the first part, Eq.(\ref{Fx in H_nm}) is further written as
\begin{equation}
F_{x}=\frac{1}{4 \pi^{2} \rho_{0} k^{2} c^{2}}  \operatorname{Re} \left\{ \sum_{n=0}^{\infty} \sum_{m=-n}^{n} C_{n} b_{n+1}^{m} \left(- H_{n,- m} H_{n+1, -m-1}^*
+ H_{n m} H_{n+1, m+1}^*
 \right) \right\}. \label{Fx in H_nm SB}
\end{equation}
which is Eq.(46) in Ref. \cite{sapozhnikov2013radiation}.

The $y$-component of ARF: 
\begin{eqnarray}
\begin{aligned}
F_{y}&=\frac{\rho_{0} \Phi_{0}^{2}}{4}  \operatorname{Re} \left\{ \sum_{n=0}^{\infty} \sum_{m=-n}^{n} C_{n} \left( b_{n+1}^{-m} a_{n}^m a_{n+1}^{m-1*}  + b_{n+1}^{m} a_{n}^m a_{n+1}^{m+1*} \right) \right\}
\\
&=\frac{1}{4 \pi^{2} \rho_{0} k^{2} c^{2}} \operatorname{Re} \left\{ \sum_{n=0}^{\infty} \sum_{m=-n}^{n} C_{n} \left[ b_{n+1}^{-m} \big(i^{n-1} H_{n m}\big) \big(i^{n} H_{n+1, m-1}\big)^*
+ b_{n+1}^{m} \big(i^{n-1} H_{n m}\big) \big(i^{n} H_{n+1, m+1}\big)^* \right] \right\}
\\
&=\frac{1}{4 \pi^{2} \rho_{0} k^{2} c^{2}} \operatorname{Re} \left\{ \sum_{n=0}^{\infty} \sum_{m=-n}^{n} i C_{n} \left( -b_{n+1}^{-m} H_{n m} H_{n+1, m-1}^*
- b_{n+1}^{m} H_{n m} H_{n+1, m+1}^* \right) \right\}
\\
&=\frac{1}{4 \pi^{2} \rho_{0} k^{2} c^{2}} \operatorname{Im} \left\{ \sum_{n=0}^{\infty} \sum_{m=-n}^{n} C_{n} \left( b_{n+1}^{-m} H_{n m} H_{n+1, m-1}^*
+ b_{n+1}^{m} H_{n m} H_{n+1, m+1}^* \right) \right\}
\end{aligned}
\label{Fy in H_nm}
\end{eqnarray}
As similar as the derivation for $F_x$, taking $C_n^{m \mp 1}$ = $C_n$ for a sphere and replacing $-m$ with $m$, $F_y$ can be also written as
\begin{equation}
F_{y}=\frac{1}{4 \pi^{2} \rho_{0} k^{2} c^{2}} \operatorname{Im} \left\{ \sum_{n=0}^{\infty} \sum_{m=-n}^{n} C_{n} b_{n+1}^{m} \left(  H_{n,-m} H_{n+1, -m-1}^* +  H_{n m} H_{n+1, m+1}^* 
\right) \right\}. \label{Fy in H_nm SB}
\end{equation}
which is Eq. (47) in Ref. \cite{sapozhnikov2013radiation}.

The $z$-component of ARF:
\begin{eqnarray}
\begin{aligned}
F_{z}&=\frac{\rho_{0} \Phi_{0}^{2}}{2}  \operatorname{Im} \left\{ \sum_{n=0}^{\infty} \sum_{m=-n}^{n} C_{n}^m c_{n+1}^{m} a_{n}^m a_{n+1}^{m*} \right\} \\
&=\frac{1}{2 \pi^{2} \rho_{0} k^{2} c^{2}} \operatorname{Im} \left\{ \sum_{n=0}^{\infty} \sum_{m=-n}^{n} C_{n}^m c_{n+1}^{m} 
\big( i^{n-1} H_{n m}\big)
\big( i^{n} H_{n+1, m}\big)^*
\right\} \\
&=\frac{1}{2 \pi^{2} \rho_{0} k^{2} c^{2}} \operatorname{Im} \left\{ \sum_{n=0}^{\infty} \sum_{m=-n}^{n} (-i) C_{n}^m c_{n+1}^{m} 
H_{n m}
H_{n+1, m}^*
\right\} \\
&=-\frac{1}{2 \pi^{2} \rho_{0} k^{2} c^{2}} \operatorname{Re} \left\{ \sum_{n=0}^{\infty} \sum_{m=-n}^{n} C_{n}^m c_{n+1}^{m} 
H_{n m}
H_{n+1, m}^* \right\} 
\end{aligned}
\label{Fz in H_nm}
\end{eqnarray}
which is Eq. (48) in Ref. \cite{sapozhnikov2013radiation} by replacing $C_n^m$ with $C_n$ for a sphere.

%---------------------------------------------------------------------
\section{\label{Appendix D}  angular momentum and ladder operators}

The ladder operators $L_{\pm}$ has the relationship with the lateral components of the angular momentum operator $L_{x,y}$: $L_{\pm}=L_{x} \pm i L_{y}$ \cite{arfken2013mathematical}. The recursion relations of ladder operators $L_{\pm}$ (or axial component of angular momentum operator $L_z$) and normalized spherical harmonics are \cite{jackson1999classical}

\begin{subequations}
\begin{eqnarray}
L_{+} Y_{n}^m=\overline{b}_{n}^{-m} Y_{n}^{m+1}, \label{L+}
\\
L_{-} Y_{n}^m=\overline{b}_{n}^m Y_{n}^{m-1}, \label{L-}
\\
L_{z} Y_{n}^m=m Y_{n}^m. \label{Lz}
\end{eqnarray}
\label{L_+-z}
\end{subequations}
with $\overline{b}_{n}^m = \sqrt{(n+m)(n-m+1)}$.

%---------------------------------------------------------------------
\section{\label{Appendix E}  Detailed derivation of ART with correct index}

\subsection{\label{Appendix E1}  Detailed derivation of $T_x$}

Based on the ART formulas of Eq. (7) from Ref.\cite{gong2019reversals}, the expression of $x$-component of ART is
\begin{eqnarray}
\begin{aligned}
T_{x} &=-\frac{\rho_{0} \Phi_{0}^{2}}{2 k } \iint_{S_{0}} \operatorname{Re}\left\{\sum_{n=0}^{\infty} \sum_{m=-n}^{n} \sum_{n'=0}^{\infty} \sum_{m'=-n'}^{n'} i^{n-n'} \left(a_{n}^{m*}+s_{n}^{m*} \right) s_{n'}^{m'} Y_{n}^{m*} L_{x} Y_{n'}^{m'} \sin \theta d \theta d \varphi \right\}  
\end{aligned} \label{Tx in a_nm and s_nm1}
\end{eqnarray}
With insertion of Eqs. (\ref{L+}) and (\ref{L-}) into (\ref{Tx in a_nm and s_nm1}) and since $L_x = (L_{+}+L_{-})/2$, 
 \begin{eqnarray}
\begin{aligned}
T_{x} =& -\frac{\rho_{0} \Phi_{0}^{2}}{4 k }  \operatorname{Re}\left\{\sum_{n=0}^{\infty} \sum_{m=-n}^{n} \sum_{n'=0}^{\infty} \sum_{m'=-n'}^{n'} i^{n-n'} \left(a_{n}^{m*}+s_{n}^{m*} \right) s_{n'}^{m'} \iint_{S_{0}} Y_{n}^{m*} (L_{+}+L_{-}) Y_{n'}^{m'} \sin \theta d \theta d \varphi \right\} 
\\
=& -\frac{\rho_{0} \Phi_{0}^{2}}{4 k }  \operatorname{Re}\left\{\sum_{n=0}^{\infty} \sum_{m=-n}^{n} \sum_{n'=0}^{\infty} \sum_{m'=-n'}^{n'-1} i^{n-n'} \left(a_{n}^{m*}+s_{n}^{m*} \right) s_{n'}^{m'} \iint_{S_{0}} Y_{n}^{m*} \bar{b}_{n'}^{-m'} Y_{n'}^{m'+1} \sin \theta d \theta d \varphi \right. \\
&+ \left. \sum_{n=0}^{\infty} \sum_{m=-n}^{n} \sum_{n'=0}^{\infty} \sum_{m'=-n'+1}^{n'} i^{n-n'} \left(a_{n}^{m*}+s_{n}^{m*} \right) s_{n'}^{m'} \iint_{S_{0}} Y_{n}^{m*} \bar{b}_{n'}^{m'} Y_{n'}^{m'-1} \sin \theta d \theta d \varphi
\right\}
\end{aligned} \label{Tx in a_nm and s_nm2}
\end{eqnarray}
The regime of $(n',m',)$in the summation symbol is listed in Table \ref{table3}.
\begin{table}[!htbp]
\small
  \caption{Regime of $(n',m')$ in normalized spherical harmonics for derivation of $F_x$ and $F_y$. Note that based on the definition in Eq. (\ref{incident potential MEM}), we have $n' \in [0,\infty]$ and $m' \in [-n',n']$.}
  \label{table3}
  \begin{tabular}{c |c | c | c}
\hline
 &  $n'$ &  $m'$ & Intersection \\
\hline 
$Y_{n'}^{m'+1}$ & $n' \in [0,\infty]$ & $ m' \in [-n'-1, n'-1]$ & $n' \in [0,\infty]$, $ m' \in [-n', n'-1]$\\

$Y_{n'}^{m'-1}$ & $n' \in [0,\infty]$ & $ m' \in [-n'+1, n'+1]$ & $n' \in [0,\infty]$, $ m' \in [-n'+1, n']$\\

 \hline
  \end{tabular}
\end{table}

Using Eq. (\ref{Ynm Ynm}), the expression of $T_x$ is
\begin{eqnarray}
\begin{aligned}
T_{x} =& -\frac{\rho_{0} \Phi_{0}^{2}}{4 k }  \operatorname{Re}\left\{\sum_{n=0}^{\infty} \sum_{m=-n}^{n} \sum_{n'=0}^{\infty} \sum_{m'=-n'}^{n'-1} i^{n-n'} \left(a_{n}^{m*}+s_{n}^{m*} \right) s_{n'}^{m'} \bar{b}_{n'}^{-m'} \delta_{nn'} \delta_{m,m'+1}  \right. \\
&+ \left. \sum_{n=0}^{\infty} \sum_{m=-n}^{n} \sum_{n'=0}^{\infty} \sum_{m'=-n'+1}^{n'} i^{n-n'} \left(a_{n}^{m*}+s_{n}^{m*} \right) s_{n'}^{m'} \bar{b}_{n'}^{m'} \delta_{nn'} \delta_{m,m'-1} \right\}
\\
=& -\frac{\rho_{0} \Phi_{0}^{2}}{4 k }  \operatorname{Re}\left\{  \sum_{n=0}^{\infty} \sum_{m=-n+1}^{n} \left(a_{n}^{m*}+s_{n}^{m*} \right) s_{n}^{m-1} \bar{b}_{n}^{-m+1} \right. \\
&+ \left. \sum_{n=0}^{\infty} \sum_{m=-n}^{n-1} \left(a_{n}^{m*}+s_{n}^{m*} \right) s_{n}^{m+1} \bar{b}_{n}^{m+1} \right\}
\end{aligned} \label{Tx in a_nm and s_nm3}
\end{eqnarray}

A re-index is necessary for the second part of Eq. (\ref{Tx in a_nm and s_nm3}) by using $q=m+1 \in [-n+1,n]$, and note that $\bar{b}_{n}^{-m+1} = \bar{b}_{n}^{m}$, we have

\begin{eqnarray}
\begin{aligned}
T_{x} =& -\frac{\rho_{0} \Phi_{0}^{2}}{4 k }  \operatorname{Re}\left\{  \sum_{n=0}^{\infty} \sum_{m=-n+1}^{n} \left(a_{n}^{m*}+s_{n}^{m*} \right) s_{n}^{m-1} \bar{b}_{n}^{-m+1} + \sum_{n=0}^{\infty} \sum_{q=-n+1}^{n} \left(a_{n}^{q-1*}+s_{n}^{q-1*} \right) s_{n}^{q} \bar{b}_{n}^{q} \right\}
\\
=& -\frac{\rho_{0} \Phi_{0}^{2}}{4 k }  \operatorname{Re}\left\{  \sum_{n=0}^{\infty} \sum_{m=-n+1}^{n} \bar{b}_{n}^{m} \Big[ \left(a_{n}^{m*}+s_{n}^{m*} \right) s_{n}^{m-1} + \left(a_{n}^{m-1*}+s_{n}^{m-1*} \right) s_{n}^{m}  \Big] \right\}
\end{aligned} \label{Tx in a_nm and s_nm4}
\end{eqnarray}
which is Eq. (\ref{MEM_x anm snm}) in Sec. \ref{sec:equivalence ART}.
%----------------------
\subsection{\label{Appendix E2}  Derivation of $T_y$}

The expression of $y$-component of ART is
\begin{eqnarray}
\begin{aligned}
T_{y} &=-\frac{\rho_{0} \Phi_{0}^{2}}{2 k } \iint_{S_{0}} \operatorname{Re}\left\{\sum_{n=0}^{\infty} \sum_{m=-n}^{n} \sum_{n'=0}^{\infty} \sum_{m'=-n'}^{n'} i^{n-n'} \left(a_{n}^{m*}+s_{n}^{m*} \right) s_{n'}^{m'} Y_{n}^{m*} L_{y} Y_{n'}^{m'} \sin \theta d \theta d \varphi \right\}  
\end{aligned} \label{Ty in a_nm and s_nm1}
\end{eqnarray}
As similar as the derivation for $T_x$, the final expression of $T_y$ in terms of $a_n^m$ and $s_n^m$ can be obtained by using Eqs. (\ref{L+}) and (\ref{L-}) into (\ref{Ty in a_nm and s_nm1}) and $L_y = (L_{+}-L_{-})/{2i}$ instaed of $L_x$, as given in Eq. (\ref{MEM_y anm snm}) and omitted here for brevity.

%---------------------------------------------------------------
\subsection{\label{Appendix E3}  Detailed derivation of $T_z$}

The expression of $z$-component of ART is
\begin{eqnarray}
\begin{aligned}
T_{z} &=-\frac{\rho_{0} \Phi_{0}^{2}}{2 k } \iint_{S_{0}} \operatorname{Re}\left\{\sum_{n=0}^{\infty} \sum_{m=-n}^{n} \sum_{n'=0}^{\infty} \sum_{m'=-n'}^{n'} i^{n-n'} \left(a_{n}^{m*}+s_{n}^{m*} \right) s_{n'}^{m'} Y_{n}^{m*} L_{z} Y_{n'}^{m'} \sin \theta d \theta d \varphi \right\}  
\end{aligned} \label{Tz in a_nm and s_nm1}
\end{eqnarray}
Insertion of Eq. (\ref{Lz}) into (\ref{Tz in a_nm and s_nm1}), we have
\begin{eqnarray}
\begin{aligned}
T_{z} &=-\frac{\rho_{0} \Phi_{0}^{2}}{2 k }  \operatorname{Re}\left\{\sum_{n=0}^{\infty} \sum_{m=-n}^{n} \sum_{n'=0}^{\infty} \sum_{m'=-n'}^{n'} i^{n-n'} \left(a_{n}^{m*}+s_{n}^{m*} \right) s_{n'}^{m'} \iint_{S_{0}} Y_{n}^{m*} m Y_{n'}^{m'} \sin \theta d \theta d \varphi \right\}  
\end{aligned} \label{Tz in a_nm and s_nm2}
\end{eqnarray}
By substituting Eq. (\ref{Ynm Ynm}) into (\ref{Tz in a_nm and s_nm2}), the final expression of $T_z$ in terms of $a_n^m$ and $s_n^m$ can be derived as
\begin{eqnarray}
\begin{aligned}
T_{z} &=-\frac{\rho_{0} \Phi_{0}^{2}}{2 k }  \operatorname{Re}\left\{\sum_{n=0}^{\infty} \sum_{m=-n}^{n} \sum_{n'=0}^{\infty} \sum_{m'=-n'}^{n'} i^{n-n'} \left(a_{n}^{m*}+s_{n}^{m*} \right) s_{n'}^{m'} m \delta_{nn'} \delta_{m m'} \right\}  
\\
&=-\frac{\rho_{0} \Phi_{0}^{2}}{2 k }  \operatorname{Re}\left\{\sum_{n=0}^{\infty} \sum_{m=-n}^{n} m \left(a_{n}^{m*}+s_{n}^{m*} \right) s_{n}^{m}   \right\}  
\end{aligned} \label{Tz in a_nm and s_nm3}
\end{eqnarray}
which is Eq. (\ref{MEM_z}) in Sec. \ref{sec:equivalence ART}.

\section{\label{Appendix F}  Equivalence of Eq. (14) and formulas by Gong \& Baudoin}
By substituting Eq. (\ref{a_nm & H_nm}) into (\ref{MEM Torque}), we can prove that the three components of ART formulas are equivalent to those in terms of $H_{nm}$ by Gong \& Baudoin [see Eqs. (10-12) in Ref. \cite{gong2020ART}], respectively. The detailed derivations are given below.

The $x$-component of ART: 
\begin{eqnarray}
\begin{aligned}
T_{x} &=-\frac{\rho_{0} \Phi_{0}^{2}}{4 k} \operatorname{Re}\left\{\sum_{n=0}^{\infty} \sum_{m=-n+1}^{n}
\overline{b}_{n}^m \overline{C}_{n}^m a_{n}^{m*} a_{n}^{m-1}\right\}
\\
&=-\frac{1}{4 \pi^{2} \rho_{0} k^{3} c^{2}} \operatorname{Re}\left\{\sum_{n=0}^{\infty} \sum_{m=-n+1}^{n}
\overline{b}_{n}^m \overline{C}_{n}^m 
\big( i^{n-1} H_{n m}\big)^*
\big( i^{n-1} H_{n, m-1}\big)
\right\}
\\
&=-\frac{1}{4 \pi^{2} \rho_{0} k^{3} c^{2}} \operatorname{Re}\left\{\sum_{n=0}^{\infty} \sum_{m=-n+1}^{n}
\overline{b}_{n}^m \overline{C}_{n}^m 
H_{n m}^* H_{n, m-1} \right\}
\end{aligned} \label{Tx in Hnm}
\end{eqnarray}
which is Eq. (10) in Ref. \cite{gong2020ART}.

The $y$-component of ART: 
\begin{eqnarray}
\begin{aligned}
T_{y} &=-\frac{\rho_{0} \Phi_{0}^{2}}{4 k} \operatorname{Im} \left\{\sum_{n=0}^{\infty} \sum_{m=-n+1}^{n} \overline{b}_{n}^m \overline{C}_{n}^m a_{n}^{ m*} a_{n}^{m-1}\right\}
\\
&=-\frac{1}{4 \pi^{2} \rho_{0} k^{3} c^{2}} \operatorname{Im}\left\{\sum_{n=0}^{\infty} \sum_{m=-n+1}^{n}
\overline{b}_{n}^m \overline{C}_{n}^m 
\big( i^{n-1} H_{n m}\big)^*
\big( i^{n-1} H_{n, m-1}\big)
\right\}
\\
&=-\frac{1}{4 \pi^{2} \rho_{0} k^{3} c^{2}} \operatorname{Im}\left\{\sum_{n=0}^{\infty} \sum_{m=-n+1}^{n}
\overline{b}_{n}^m \overline{C}_{n}^m 
H_{n m}^* H_{n, m-1} \right\}
\end{aligned} \label{Ty in Hnm}
\end{eqnarray}
which is Eq. (11) in Ref. \cite{gong2020ART}.

The $z$-component of ART: 
\begin{eqnarray}
\begin{aligned}
T_{z} &=-\frac{\rho_{0} \Phi_{0}^{2}}{2 k }  \operatorname{Re}\left\{\sum_{n=0}^{\infty} \sum_{m=-n}^{n} m \overline{D}_{n}^m a_{n }^{m*} a_{n}^m   \right\}  
\\
&=-\frac{1}{2 \pi^{2} \rho_{0} k^{3} c^{2}} \operatorname{Re}\left\{\sum_{n=0}^{\infty} \sum_{m=-n}^{n}
m \overline{D}_{n}^m
\big( i^{n-1} H_{n m}\big)^*
\big( i^{n-1} H_{n m}\big)
\right\}
\\
&=-\frac{1}{2 \pi^{2} \rho_{0} k^{3} c^{2}} \operatorname{Re}\left\{\sum_{n=0}^{\infty} \sum_{m=-n}^{n}
m \overline{D}_{n}^m
H_{n m}^* H_{n m}
\right\}
\end{aligned} \label{Tz in Hnm}
\end{eqnarray}
which is Eq. (12) in Ref. \cite{gong2020ART}.

%--------------------------------------
\bibliography{main}

\begin{thebibliography}{10}
\def\enquote#1,{``#1,''}
\def\enxquote#1{``#1''}
\expandafter\ifx\csname url\endcsname\relax
  \def\url#1{\texttt{#1}}\fi
\expandafter\ifx\csname urlprefix\endcsname\relax\def\urlprefix{URL }\fi
\providecommand{\bibinfo}[2]{#2}
\def\plainquote#1{``#1''}
\providecommand{\noopsort}[1]{}
\providecommand{\switchargs}[2]{#2#1}
\providecommand{\dourl}[1]{\href{http://#1}{\nolinkurl{#1}}}
  \def\eatspace #1{#1}

\bibitem{pm_rayleigh_1902}
\bibinfo{author}{L.~Rayleigh}, \enquote{\bibinfo{title}{On the pressure of
  vibration}},  \bibinfo{journal}{Philos. Mag.} \textbf{3},
  \bibinfo{pages}{338--346} (\bibinfo{year}{1902}).

\bibitem{pm_rayleigh_1905}
\bibinfo{author}{L.~Rayleigh}, \enquote{\bibinfo{title}{On the momentum and
  pressure of gaseous vibrations, and on the connection with the virial
  theorem}},  \bibinfo{journal}{Philos. Mag.} \textbf{10},
  \bibinfo{pages}{366--374} (\bibinfo{year}{1905}).

\bibitem{ra_biquard_1932a}
\bibinfo{author}{P.~Biquard}, \enquote{\bibinfo{title}{Les ondes
  ultra-sonores}},  \bibinfo{journal}{Rev. D'Acous.} \textbf{1},
  \bibinfo{pages}{93--109} (\bibinfo{year}{1932}).

\bibitem{ra_biquard_1932b}
\bibinfo{author}{P.~Biquard}, \enquote{\bibinfo{title}{Les ondes ultra-sonores
  ii}},  \bibinfo{journal}{Rev. D'Acous.} \textbf{1}, \bibinfo{pages}{315--355}
  (\bibinfo{year}{1932}).

\bibitem{jpr_brillouin_1925}
\bibinfo{author}{L.~Brillouin}, \enquote{\bibinfo{title}{Les tensions de
  radiation; leur interprétation en mécanique classique et en relativité}},
  \bibinfo{journal}{J. Phys. Radium} \textbf{6}, \bibinfo{pages}{337--353}
  (\bibinfo{year}{1925}).

\bibitem{ap_brillouin_1925}
\bibinfo{author}{L.~Brillouin}, \enquote{\bibinfo{title}{Sur les tensions de
  radiation}},  \bibinfo{journal}{Ann. Phys.} \textbf{4},
  \bibinfo{pages}{528--86} (\bibinfo{year}{1925}).

\bibitem{king1934acoustic}
\bibinfo{author}{L.~V. King}, \enquote{\bibinfo{title}{On the acoustic
  radiation pressure on spheres}},  \bibinfo{journal}{Proc. R. Soc. London}
  \textbf{147}(861), \bibinfo{pages}{212--240} (\bibinfo{year}{1934}).

\bibitem{a_yosika_1955}
\bibinfo{author}{K.~Yosika} and \bibinfo{author}{Y.~Kawasima},
  \enquote{\bibinfo{title}{Acoustic radiation pressure on a compressible
  sphere}},  \bibinfo{journal}{Acustica} \textbf{5}, \bibinfo{pages}{167--173}
  (\bibinfo{year}{1955}).

\bibitem{jasa_hasegawa_1969}
\bibinfo{author}{T.~Hasegawa} and \bibinfo{author}{K.~Yosika},
  \enquote{\bibinfo{title}{Acoustic radiation pressure on a solide elastic
  sphere}},  \bibinfo{journal}{J. Acoust. Soc. Am.} \textbf{46},
  \bibinfo{pages}{1119--1143} (\bibinfo{year}{1969}).

\bibitem{jasa_embleton_1954}
\bibinfo{author}{T.~Embleton}, \enquote{\bibinfo{title}{Mean force on a sphere
  in a spherical sound field}},  \bibinfo{journal}{J. Acoust. Soc. Am.}
  \textbf{26}, \bibinfo{pages}{40--45} (\bibinfo{year}{1954}).

\bibitem{jasa_chen_1996}
\bibinfo{author}{X.~Chen} and \bibinfo{author}{R.~Apfel},
  \enquote{\bibinfo{title}{Radiation force on a spherical object in the field
  of a focused cylindrical transducer}},  \bibinfo{journal}{J. Acoust. Soc.
  Am.} \textbf{101}, \bibinfo{pages}{2443--2447} (\bibinfo{year}{1996}).

\bibitem{baudoin2020acoustic}
\bibinfo{author}{M.~Baudoin} and \bibinfo{author}{J.-L. Thomas},
  \enquote{\bibinfo{title}{Acoustic tweezers for particle and fluid
  micromanipulation}},  \bibinfo{journal}{Annu. Rev. Fluid Mech.} \textbf{52},
  \bibinfo{pages}{205--234} (\bibinfo{year}{2020}).

\bibitem{spd_gorkov_1962}
\bibinfo{author}{L.~Gor'ov}, \enquote{\bibinfo{title}{On the forces acting on a
  small particle in an acoustic field in an ideal fluid}},
  \bibinfo{journal}{Sov. Phys. Dokl} \textbf{6}, \bibinfo{pages}{773--775}
  (\bibinfo{year}{1962}).

\bibitem{jasa_busse_1981}
\bibinfo{author}{F.~H. Busse} and \bibinfo{author}{T.~G. Wang},
  \enquote{\bibinfo{title}{Torque generated by orthogonal acoustic
  waves—theory}},  \bibinfo{journal}{J. Acoust. Soc. Am.} \textbf{69}(6),
  \bibinfo{pages}{1634--1638} (\bibinfo{year}{1981}).

\bibitem{zhang2011angular}
\bibinfo{author}{L.~Zhang} and \bibinfo{author}{P.~L. Marston},
  \enquote{\bibinfo{title}{Angular momentum flux of nonparaxial acoustic vortex
  beams and torques on axisymmetric objects}},  \bibinfo{journal}{Phys. Rev. E}
  \textbf{84}(6), \bibinfo{pages}{065601} (\bibinfo{year}{2011}).

\bibitem{sapozhnikov2013radiation}
\bibinfo{author}{O.~A. Sapozhnikov} and \bibinfo{author}{M.~R. Bailey},
  \enquote{\bibinfo{title}{Radiation force of an arbitrary acoustic beam on an
  elastic sphere in a fluid}},  \bibinfo{journal}{J. Acoust. Soc. Am.}
  \textbf{133}(2), \bibinfo{pages}{661--676} (\bibinfo{year}{2013}).

\bibitem{gong2020ART}
\bibinfo{author}{Z.~Gong} and \bibinfo{author}{M.~Baudoin},
  \enquote{\bibinfo{title}{Radiation torque on a particle in a fluid: An
  angular spectrum based compact expression}},  \bibinfo{journal}{J. Acoust.
  Soc. Am.} \textbf{148}(5), \bibinfo{pages}{3131--3140}
  (\bibinfo{year}{2020}).

\bibitem{silva2011expression}
\bibinfo{author}{G.~T. Silva}, \enquote{\bibinfo{title}{An expression for the
  radiation force exerted by an acoustic beam with arbitrary wavefront (l)}},
  \bibinfo{journal}{J. Acoust. Soc. Am.} \textbf{130}(6),
  \bibinfo{pages}{3541--3544} (\bibinfo{year}{2011}).

\bibitem{silva2012radiation}
\bibinfo{author}{G.~Silva}, \bibinfo{author}{T.~Lobo}, and
  \bibinfo{author}{F.~Mitri}, \enquote{\bibinfo{title}{Radiation torque
  produced by an arbitrary acoustic wave}},  \bibinfo{journal}{EPL (Europhysics
  Letters)} \textbf{97}(5), \bibinfo{pages}{54003} (\bibinfo{year}{2012}).

\bibitem{baresch2013three}
\bibinfo{author}{D.~Baresch}, \bibinfo{author}{J.-L. Thomas}, and
  \bibinfo{author}{R.~Marchiano}, \enquote{\bibinfo{title}{Three-dimensional
  acoustic radiation force on an arbitrarily located elastic sphere}},
  \bibinfo{journal}{J. Acoust. Soc. Am.} \textbf{133}(1),
  \bibinfo{pages}{25--36} (\bibinfo{year}{2013}).

\bibitem{gong2019t}
\bibinfo{author}{Z.~Gong}, \bibinfo{author}{P.~L. Marston}, and
  \bibinfo{author}{W.~Li}, \enquote{\bibinfo{title}{T-matrix evaluation of
  three-dimensional acoustic radiation forces on nonspherical objects in bessel
  beams with arbitrary order and location}},  \bibinfo{journal}{Phys. Rev. E}
  \textbf{99}(6), \bibinfo{pages}{063004} (\bibinfo{year}{2019}).

\bibitem{gong2019reversals}
\bibinfo{author}{Z.~Gong}, \bibinfo{author}{P.~L. Marston}, and
  \bibinfo{author}{W.~Li}, \enquote{\bibinfo{title}{Reversals of acoustic
  radiation torque in bessel beams using theoretical and numerical
  implementations in three dimensions}},  \bibinfo{journal}{Phys. Rev. Applied}
  \textbf{11}(6), \bibinfo{pages}{064022} (\bibinfo{year}{2019}).

\bibitem{zhao2019computation}
\bibinfo{author}{D.~Zhao}, \bibinfo{author}{J.-L. Thomas}, and
  \bibinfo{author}{R.~Marchiano}, \enquote{\bibinfo{title}{Computation of the
  radiation force exerted by the acoustic tweezers using pressure field
  measurements}},  \bibinfo{journal}{J. Acoust. Soc. Am.} \textbf{146}(3),
  \bibinfo{pages}{1650--1660} (\bibinfo{year}{2019}).

\bibitem{westervelt1951theory}
\bibinfo{author}{P.~J. Westervelt}, \enquote{\bibinfo{title}{The theory of
  steady forces caused by sound waves}},  \bibinfo{journal}{J. Acoust. Soc.
  Am.} \textbf{23}(3), \bibinfo{pages}{312--315} (\bibinfo{year}{1951}).

\bibitem{westervelt1957acoustic}
\bibinfo{author}{P.~J. Westervelt}, \enquote{\bibinfo{title}{Acoustic radiation
  pressure}},  \bibinfo{journal}{J. Acoust. Soc. Am.} \textbf{29}(1),
  \bibinfo{pages}{26--29} (\bibinfo{year}{1957}).

\bibitem{maidanik1958torques}
\bibinfo{author}{G.~Maidanik}, \enquote{\bibinfo{title}{Torques due to
  acoustical radiation pressure}},  \bibinfo{journal}{J. Acoust. Soc. Am.}
  \textbf{30}(7), \bibinfo{pages}{620--623} (\bibinfo{year}{1958}).

\bibitem{jasa_fan_2008}
\bibinfo{author}{Z.~Fan}, \bibinfo{author}{D.~Mei}, \bibinfo{author}{K.~Yang},
  and \bibinfo{author}{Z.~Chen}, \enquote{\bibinfo{title}{Acoustic radiation
  torque on an irregularly shaped scattered in an arbitrary sound field}},
  \bibinfo{journal}{J. Acoust. Soc. Am.} \textbf{124},
  \bibinfo{pages}{2727--2732} (\bibinfo{year}{2008}).

\bibitem{jasa_zhang_2011}
\bibinfo{author}{L.~Zhang} and \bibinfo{author}{P.~Marston},
  \enquote{\bibinfo{title}{Acoustic radiation torque and the conservation of
  angular momentum (l)}},  \bibinfo{journal}{J. Acoust. Soc. Am.}
  \textbf{129}(4), \bibinfo{pages}{1679--1680} (\bibinfo{year}{2011}).

\bibitem{jasj_hasegawa_2000}
\bibinfo{author}{T.~Hasegawa}, \bibinfo{author}{T.~Kido},
  \bibinfo{author}{T.~Iusizka}, and \bibinfo{author}{C.~Matsuoak},
  \enquote{\bibinfo{title}{A general theory of rayleigh and langevin radiation
  pressures}},  \bibinfo{journal}{J. Acoust. Soc. Am.} \textbf{21},
  \bibinfo{pages}{145--152} (\bibinfo{year}{2000}).

\bibitem{thomas2017acoustical}
\bibinfo{author}{J.-L. Thomas}, \bibinfo{author}{R.~Marchiano}, and
  \bibinfo{author}{D.~Baresch}, \enquote{\bibinfo{title}{Acoustical and optical
  radiation pressure and the development of single beam acoustical tweezers}},
  \bibinfo{journal}{J. Quant. Spectrosc. Radiat. Transf.} \textbf{195},
  \bibinfo{pages}{55--65} (\bibinfo{year}{2017}).

\bibitem{pre_jimenez_2016}
\bibinfo{author}{N.~Jimenez}, \bibinfo{author}{R.~Pico},
  \bibinfo{author}{V.~Sanchez-Morcillo}, \bibinfo{author}{V.~Romero-Garcia},
  \bibinfo{author}{L.~Garcia-Raffi}, and \bibinfo{author}{K.~Staliunas},
  \enquote{\bibinfo{title}{Formation of high-order acoustic bessel beams by
  spiral diffraction gratings}},  \bibinfo{journal}{Phys. Rev. E}
  \textbf{94}(5), \bibinfo{pages}{053004} (\bibinfo{year}{2016}).

\bibitem{prap_riaud_2017}
\bibinfo{author}{A.~Riaud}, \bibinfo{author}{M.~Baudoin},
  \bibinfo{author}{O.~Bou~Matar}, \bibinfo{author}{L.~Becerra}, and
  \bibinfo{author}{J.-L. Thomas}, \enquote{\bibinfo{title}{Selective
  manipulation of microscopic particles with precursors swirling rayleigh
  waves}},  \bibinfo{journal}{Phys. Rev. Applied} \textbf{7},
  \bibinfo{pages}{024007} (\bibinfo{year}{2017}).

\bibitem{apl_jimenez_2018}
\bibinfo{author}{N.~Jimenez}, \bibinfo{author}{V.~Romero-Garcia},
  \bibinfo{author}{L.~Garcia-Raffi}, \bibinfo{author}{F.~Camarena}, and
  \bibinfo{author}{K.~Staliunas}, \enquote{\bibinfo{title}{Sharp acoustic
  vortex focusing by fresnel-spiral-zone plates}},  \bibinfo{journal}{Appl.
  Phys. Lett.} \textbf{112}(20), \bibinfo{pages}{204101}
  (\bibinfo{year}{2018}).

\bibitem{sa_baudoin_2019}
\bibinfo{author}{M.~Baudoin}, \bibinfo{author}{J.-C. Gerbedoen},
  \bibinfo{author}{A.~Riaud}, \bibinfo{author}{O.~Bou~Matar},
  \bibinfo{author}{N.~Smagin}, and \bibinfo{author}{J.-L. Thomas},
  \enquote{\bibinfo{title}{Folding a focalized acoustical vortex on a flat
  holographic transducer: miniaturized selective acoustical tweezers}},
  \bibinfo{journal}{Sci. Adv.} \textbf{5}, \bibinfo{pages}{eaav1967}
  (\bibinfo{year}{2019}).

\bibitem{baudoin2020naturecell}
\bibinfo{author}{M.~Baudoin}, \bibinfo{author}{J.-L. Thomas},
  \bibinfo{author}{R.~A. Sahely}, \bibinfo{author}{J.-C. Gerbedoen},
  \bibinfo{author}{Z.~Gong}, \bibinfo{author}{A.~Sivery},
  \bibinfo{author}{O.~Matar}, \bibinfo{author}{N.~Smagin},
  \bibinfo{author}{P.~Favreau}, and \bibinfo{author}{A.~Vlandas},
  \enquote{\bibinfo{title}{Spatially selective manipulation of cells with
  single-beam acoustical tweezers}},  \bibinfo{journal}{Nat. Commu.}
  \textbf{11}, \bibinfo{pages}{4244} (\bibinfo{year}{2020}).

\bibitem{gong2018thesis}
\bibinfo{author}{Z.~Gong}, \enquote{\bibinfo{title}{Study on acoustic
  scattering characteristics of objects in {B}essel beams and the related
  radiation force and torque}}, \bibinfo{type}{Ph.{D}. dissertation},
  \bibinfo{school}{Huazhong University of Science and Technology},
  \bibinfo{address}{Wuhan, China}, \bibinfo{year}{2018}.

\bibitem{gong2016arbitrary}
\bibinfo{author}{Z.~Gong}, \bibinfo{author}{W.~Li}, \bibinfo{author}{F.~G.
  Mitri}, \bibinfo{author}{Y.~Chai}, and \bibinfo{author}{Y.~Zhao},
  \enquote{\bibinfo{title}{Arbitrary scattering of an acoustical bessel beam by
  a rigid spheroid with large aspect-ratio}},  \bibinfo{journal}{Journal of
  Sound and Vibration} \textbf{383}, \bibinfo{pages}{233--247}
  (\bibinfo{year}{2016}).

\bibitem{gong2017t}
\bibinfo{author}{Z.~Gong}, \bibinfo{author}{W.~Li}, \bibinfo{author}{Y.~Chai},
  \bibinfo{author}{Y.~Zhao}, and \bibinfo{author}{F.~G. Mitri},
  \enquote{\bibinfo{title}{T-matrix method for acoustical bessel beam
  scattering from a rigid finite cylinder with spheroidal endcaps}},
  \bibinfo{journal}{Ocean Engineering} \textbf{129}, \bibinfo{pages}{507--519}
  (\bibinfo{year}{2017}).

\bibitem{gong2017analysis}
\bibinfo{author}{W.~Li}, \bibinfo{author}{Y.~Chai}, \bibinfo{author}{Z.~Gong},
  and \bibinfo{author}{P.~L. Marston}, \enquote{\bibinfo{title}{Analysis of
  forward scattering of an acoustical zeroth-order bessel beam from rigid
  complicated (aspherical) structures}},  \bibinfo{journal}{J. Quant.
  Spectrosc. Radiat. Transf.} \textbf{200}, \bibinfo{pages}{146--162}
  (\bibinfo{year}{2017}).

\bibitem{marston2006axial}
\bibinfo{author}{P.~L. Marston}, \enquote{\bibinfo{title}{Axial radiation force
  of a bessel beam on a sphere and direction reversal of the force}},
  \bibinfo{journal}{J. Acoust. Soc. Am.} \textbf{120}(6),
  \bibinfo{pages}{3518--3524} (\bibinfo{year}{2006}).

\bibitem{arfken2013mathematical}
\bibinfo{author}{G.~B. Arfken}, \bibinfo{author}{H.~J. Weber}, and
  \bibinfo{author}{F.~E. Harris}, \emph{\bibinfo{title}{Mathematical methods
  for physicists}}, \bibinfo{}{7th} ed.  (\bibinfo{publisher}{Academic Press},
  \bibinfo{address}{New York}, \bibinfo{year}{2013}), pp.
  \bibinfo{pages}{756--765}.

\bibitem{gong2017multipole}
\bibinfo{author}{Z.~Gong}, \bibinfo{author}{P.~L. Marston},
  \bibinfo{author}{W.~Li}, and \bibinfo{author}{Y.~Chai},
  \enquote{\bibinfo{title}{Multipole expansion of acoustical bessel beams with
  arbitrary order and location}},  \bibinfo{journal}{J. Acoust. Soc. Am.}
  \textbf{141}(6), \bibinfo{pages}{EL574--EL578} (\bibinfo{year}{2017}).

\bibitem{silva2011off}
\bibinfo{author}{G.~T. Silva}, \enquote{\bibinfo{title}{Off-axis scattering of
  an ultrasound bessel beam by a sphere}},  \bibinfo{journal}{IEEE Trans.
  Ultrason. Ferroelectr. Freq. Control} \textbf{58}(2),
  \bibinfo{pages}{298--304} (\bibinfo{year}{2011}).

\bibitem{gong2019resonance}
\bibinfo{author}{W.~Li}, \bibinfo{author}{Q.~Gui}, and
  \bibinfo{author}{Z.~Gong}, \enquote{\bibinfo{title}{Resonance scattering of
  an arbitrary bessel beam by a spherical object}},  \bibinfo{journal}{IEEE
  Trans. Ultrason. Ferroelectr. Freq. Control} \textbf{66}(8),
  \bibinfo{pages}{1364--1372} (\bibinfo{year}{2019}).

\bibitem{silva2018acoustic}
\bibinfo{author}{G.~T. Silva} and \bibinfo{author}{B.~W. Drinkwater},
  \enquote{\bibinfo{title}{Acoustic radiation force exerted on a small
  spheroidal rigid particle by a beam of arbitrary wavefront: Examples of
  traveling and standing plane waves}},  \bibinfo{journal}{J. Acoust. Soc. Am.}
  \textbf{144}(5), \bibinfo{pages}{EL453--EL459} (\bibinfo{year}{2018}).

\bibitem{silva2020nonlinear}
\bibinfo{author}{E.~B. Lima}, \bibinfo{author}{J.~P. Leao-Neto},
  \bibinfo{author}{A.~S. Marques}, \bibinfo{author}{G.~C. Silva},
  \bibinfo{author}{J.~H. Lopes}, and \bibinfo{author}{G.~T. Silva},
  \enquote{\bibinfo{title}{Nonlinear interaction of acoustic waves with a
  spheroidal particle: radiation force and torque effects}},
  \bibinfo{journal}{Phys. Rev. Applied} \textbf{13}, \bibinfo{pages}{064048}
  (\bibinfo{year}{2020}).

\bibitem{baresch2018orbital}
\bibinfo{author}{D.~Baresch}, \bibinfo{author}{J.-L. Thomas}, and
  \bibinfo{author}{R.~Marchiano}, \enquote{\bibinfo{title}{Orbital angular
  momentum transfer to stably trapped elastic particles in acoustical vortex
  beams}},  \bibinfo{journal}{Phys. Rev. Lett.} \textbf{121}(7),
  \bibinfo{pages}{074301} (\bibinfo{year}{2018}).

\bibitem{bostrom1980multiple}
\bibinfo{author}{A.~Bostr{\"o}m}, \enquote{\bibinfo{title}{Multiple scattering
  of elastic waves by bounded obstacles}},  \bibinfo{journal}{J. Acoust. Soc.
  Am.} \textbf{67}(2), \bibinfo{pages}{399--413} (\bibinfo{year}{1980}).

\bibitem{silva2014acoustic}
\bibinfo{author}{G.~T. Silva} and \bibinfo{author}{H.~Bruus},
  \enquote{\bibinfo{title}{Acoustic interaction forces between small particles
  in an ideal fluid}},  \bibinfo{journal}{Phys. Rev. E} \textbf{90}(6),
  \bibinfo{pages}{063007} (\bibinfo{year}{2014}).

\bibitem{gong2019particle}
\bibinfo{author}{Z.~Gong} and \bibinfo{author}{M.~Baudoin},
  \enquote{\bibinfo{title}{Particle assembly with synchronized acoustic
  tweezers}},  \bibinfo{journal}{Phys. Rev. Applied} \textbf{12}(2),
  \bibinfo{pages}{024045} (\bibinfo{year}{2019}).

\bibitem{prap_gong_2020}
\bibinfo{author}{Z.~Gong} and \bibinfo{author}{M.~Baudoin},
  \enquote{\bibinfo{title}{Three-dimensional trapping and assembly of small
  particles with synchronized spherical acoustical vortices}},
  \bibinfo{journal}{Phys. Rev. Appl.} \textbf{14}, \bibinfo{pages}{064002}
  (\bibinfo{year}{2020}).

\bibitem{marston2007acoustic}
\bibinfo{author}{P.~L. Marston}, \enquote{\bibinfo{title}{Acoustic beam
  scattering and excitation of sphere resonance: Bessel beam example}},
  \bibinfo{journal}{J. Acoust. Soc. Am.} \textbf{122}(1),
  \bibinfo{pages}{247--252} (\bibinfo{year}{2007}).

\bibitem{marston2009erratum}
\bibinfo{author}{P.~L. Marston}, \enquote{\bibinfo{title}{Erratum: Acoustic
  beam scattering and excitation of sphere resonance: Bessel beam example}},
  \bibinfo{journal}{J. Acoust. Soc. Am.} \textbf{125}(6),
  \bibinfo{pages}{4092--4092} (\bibinfo{year}{2009}).

\bibitem{zhang2011geometrical}
\bibinfo{author}{L.~Zhang} and \bibinfo{author}{P.~L. Marston},
  \enquote{\bibinfo{title}{Geometrical interpretation of negative radiation
  forces of acoustical bessel beams on spheres}},  \bibinfo{journal}{Phys. Rev.
  E} \textbf{84}(3), \bibinfo{pages}{035601} (\bibinfo{year}{2011}).

\bibitem{fan2019trapping}
\bibinfo{author}{X.-D. Fan} and \bibinfo{author}{L.~Zhang},
  \enquote{\bibinfo{title}{Trapping force of acoustical bessel beams on a
  sphere and stable tractor beams}},  \bibinfo{journal}{Phys. Rev. Applied}
  \textbf{11}(1), \bibinfo{pages}{014055} (\bibinfo{year}{2019}).

\bibitem{baresch2013spherical}
\bibinfo{author}{D.~Baresch}, \bibinfo{author}{J.-L. Thomas}, and
  \bibinfo{author}{R.~Marchiano}, \enquote{\bibinfo{title}{Spherical vortex
  beams of high radial degree for enhanced single-beam tweezers}},
  \bibinfo{journal}{J. Appl. Phys.} \textbf{113}(18), \bibinfo{pages}{184901}
  (\bibinfo{year}{2013}).

\bibitem{baresch2016observation}
\bibinfo{author}{D.~Baresch}, \bibinfo{author}{J.-L. Thomas}, and
  \bibinfo{author}{R.~Marchiano}, \enquote{\bibinfo{title}{Observation of a
  single-beam gradient force acoustical trap for elastic particles: acoustical
  tweezers}},  \bibinfo{journal}{Phys. Rev. lett.} \textbf{116}(2),
  \bibinfo{pages}{024301} (\bibinfo{year}{2016}).

\bibitem{gong2018conference}
\bibinfo{author}{Z.~Gong}, \bibinfo{author}{Y.~Chai}, and
  \bibinfo{author}{W.~Li}, \enquote{\bibinfo{title}{Reversals of acoustic
  radiation force and torque in a single {B}essel beam{:} acoustic tweezers
  numerical toolbox}},  \bibinfo{journal}{Acoustofluidics2018, Lille, France}
  (\bibinfo{year}{2018}).

\bibitem{nieminen2007optical}
\bibinfo{author}{T.~A. Nieminen}, \bibinfo{author}{V.~L. Loke},
  \bibinfo{author}{A.~B. Stilgoe}, \bibinfo{author}{G.~Kn{\"o}ner},
  \bibinfo{author}{A.~M. Bra{\'n}czyk}, \bibinfo{author}{N.~R. Heckenberg}, and
  \bibinfo{author}{H.~Rubinsztein-Dunlop}, \enquote{\bibinfo{title}{Optical
  tweezers computational toolbox}},  \bibinfo{journal}{” J. Opt. A, Pure
  Appl. Opt.} \textbf{9}(8), \bibinfo{pages}{S196} (\bibinfo{year}{2007}).

\bibitem{jackson1999classical}
\bibinfo{author}{J.~D. Jackson}, \emph{\bibinfo{title}{Classical
  electrodynamics}}, \bibinfo{}{3th} ed.  (\bibinfo{publisher}{John Wiley \&
  Sons}, \bibinfo{address}{New York}, \bibinfo{year}{1999}), p.
  \bibinfo{pages}{428}.

\end{thebibliography}

\end{document}